\documentclass[12pt,a4paper]{article}
\usepackage{amsfonts}
\usepackage{comment}
\usepackage{amssymb}
\usepackage{graphicx}
\usepackage{natbib}
\usepackage{amsmath}
\usepackage{amsthm}
\usepackage{paralist}
\usepackage{subcaption}
\usepackage{xcolor}
\usepackage{multirow}
\usepackage{booktabs}
\IfFileExists{dsfont.sty}{\usepackage{dsfont}}{}
\usepackage{indentfirst}
\usepackage{enumerate}
\usepackage{longtable}
\usepackage{array}
\usepackage{caption}
\IfFileExists{bbm.sty}{\usepackage{bbm}}{}
\usepackage{placeins}
\IfFileExists{algpseudocode.sty}{\usepackage{algpseudocode}}{} 
\usepackage{lscape}
\newdimen\nodeDist
\usepackage[hyperindex,breaklinks,hypertexnames=false]{hyperref}
\hypersetup{colorlinks=true,       
	linkcolor=red,       
	citecolor=blue,        
	filecolor=magenta,      
	urlcolor=cyan           
}  

\usepackage[linesnumbered,boxed,ruled,commentsnumbered]{algorithm2e}

\usepackage[toc,page]{appendix}
\usepackage{setspace}

\usepackage{bibunits}
\defaultbibliographystyle{chicago}
\defaultbibliography{ref}

\usepackage[left = 1in, right = 1in, top = 1in, bottom = 1in]{geometry}
\usepackage{titlesec}
\titleformat*{\section}{\large\bfseries}
\titleformat*{\subsection}{\normalsize\bfseries}
\titleformat*{\subsubsection}{\normalsize\bfseries}
\titleformat*{\paragraph}{\normalsize\bfseries}

\titlespacing{\section}{0pt}{*1.6}{*1.6}
\titlespacing{\subsection}{0pt}{*1.4}{*1.4}
\titlespacing{\subsubsection}{0pt}{*1.2}{*1.2}
\titlespacing{\paragraph}{0pt}{*1.2}{*1.2}

\AtBeginDocument{%
  \setlength{\abovedisplayshortskip}{1pt}%
  \setlength{\belowdisplayshortskip}{1pt}%
  \setlength{\abovedisplayskip}{1pt}%
  \setlength{\belowdisplayskip}{1pt}%
}
  
\newtheorem{remark}{Remark}
\newtheorem{theorem}{Theorem}

\newtheorem{proposition}{Proposition}
\newtheorem{corollary}{Corollary}
\newtheorem{condition}{Condition}

\allowdisplaybreaks

\newcommand{\B}{ \textbf{B}}
\newcommand{\W}{\textbf{W}}
\newcommand{\U}{\textbf{U}}
\newcommand{\V}{\textbf{V}}

\newcommand{\Y}{\textbf{Y}}

\definecolor{revisionpurple}{RGB}{0,0,0}

\definecolor{notepurple}{HTML}{7030A0}

\usepackage{authblk}

\begin{document}

\title{\Large{\bf
A Hierarchical Group Factor Model with \\
Characteristic-Driven Global and Local Factors}
}

\author[1]{\small Jun Zhang}
\author[2]{\small Yongqiang Wu}
\author[2]{\small Dan Pu}
\author[2]{\small Wei Lan}
\author[3]{\small Guanhao Feng}

\affil[1]{\small{School of Statistics and Data Science, Southeast University}}
\affil[2]{\small{School of Statistics and Data Science, Southwestern University of Finance and Economics
}}
\affil[3]{\small{City University of Hong Kong}}

\vspace{-1cm}

\date{ }

\maketitle

\begin{abstract}
\small
\noindent
We propose a Hierarchical Group Factor Model (HGFM) for jointly estimating global and group-specific factors with characteristic-driven portfolio weights.
Both factor types are constructed from contemporaneous returns, producing endogeneity that we address by quasi-maximum likelihood estimation.
Conditional on the interaction matrices, we separate the global and local coefficient components and normalize their factor representations.
Under a fixed characteristic design and suitable regularity conditions, we establish local consistency of the coefficient estimators and asymptotic normality for fixed blocks of the normalized factor parameters as both panel dimensions diverge.
We also propose an information criterion and establish its consistency for jointly selecting the global rank and the common local rank.
In an industry-grouped application to S\&P 100 constituents, the HGFM achieves higher explanatory power for realized excess returns and a higher portfolio Sharpe ratio than benchmark asset-pricing models.

\bigskip


\end{abstract}

\newpage

\setstretch{1.8}

\section{Introduction}
How can we separate common and group-specific dependence when both global and local factors are constructed from observed responses using characteristic-driven weights? For stock returns grouped by industry, the distinction is between comovement across industries and additional comovement within each industry. Standard factor models summarize common variation in high-dimensional panels \citep[e.g.,][]{bai2002determining}, whereas group factor models (GFMs) distinguish global factors shared across groups from local factors specific to each group. Grouped heterogeneity also motivates sparse asset-pricing models \citep{cong2024sparse} and group-specific predictability of mutual fund alphas \citep{cui2025heterogeneous}. Characteristic-based portfolios formed across the full stock panel and within individual industries may capture overlapping variation in returns. This motivates jointly specifying and estimating the global and local components within a common return model. 
We develop a framework in which predetermined asset characteristics generate time-varying weights for both global and local factors.

{\citet*{andreou2019inference} develop inference for group factor models using within-group principal components and cross-group canonical correlations.} {We use their common and group-specific latent factor structure as a benchmark.}
Let $y_{s,it}$ denote the response variable for the $i$th cross-sectional unit at time $t$ in group $s$, where $s = 1, \ldots, S$, $i = 1, \ldots, N_s$, and $t = 1, \ldots, T.$ Define $\Y_{s,t} = (y_{s,1t}, \ldots,y_{s, N_s t})^\top \in \mathbb{R}^{N_s}$ as the vector of response variables for group $s$ at time $t$. The group factor model is then specified as follows: 
\begingroup\setlength{\abovedisplayskip}{4pt}\setlength{\belowdisplayskip}{4pt}\setlength{\abovedisplayshortskip}{2pt}\setlength{\belowdisplayshortskip}{2pt}
\begin{equation} \label{eq:global_model_s_intro} 
\Y_{s,t} = \B_s \mathbf f_{g,t} + \U_s \mathbf f_{l,s,t} + \boldsymbol\varepsilon_{s,t}, 
\end{equation}
\endgroup
where $\mathbf f_{g,t}\in\mathbb R^{r_g}$ and $\mathbf f_{l,s,t}\in\mathbb R^{r_l}$ are the latent global and local factors, respectively; $\B_s\in\mathbb R^{N_s\times r_g}$ and $\U_s\in\mathbb R^{N_s\times r_l}$ are their loading matrices; and $\boldsymbol\varepsilon_{s,t}\in\mathbb R^{N_s}$ is the idiosyncratic error. Thus, $\mathbf f_{g,t}$ affects all groups, whereas $\mathbf f_{l,s,t}$ affects only group $s$.

The latent-factor specification in \eqref{eq:global_model_s_intro} leaves the relation between asset characteristics and factor-portfolio weights unspecified. Empirical asset pricing often constructs observable factors by sorting securities on characteristics and forming long--short portfolios, as in the Fama--French factors \citep{Fama1993Common,fama2015five}. The number of such factors has grown large \citep{jensen2023there}, and \citet{feng2020taming} test whether a new factor adds explanatory power for expected returns beyond the existing set.

We express both types of factors in \eqref{eq:global_model_s_intro} as linear combinations of contemporaneous returns, with time-varying weights determined by characteristics, and estimate their fixed coefficients jointly.
\citet*{kelly2019characteristics} use characteristics to parameterize time-varying factor loadings; our characteristics instead determine portfolio weights:
\begingroup\setlength{\abovedisplayskip}{4pt}\setlength{\belowdisplayskip}{4pt}\setlength{\abovedisplayshortskip}{2pt}\setlength{\belowdisplayshortskip}{2pt}
\begin{equation} \label{eq:observable_factor_construction_intro}
\begin{split}
   \mathbf f_{g,t}=\W_{t-1}\Y_t,
  \mathbf f_{l,s,t}=\V_{s,t-1}\Y_{s,t}, \\
   \mathbf W_{t-1}=\mathbf C_g\mathbf Z_{t-1}, \mathbf V_{s,t-1}=\mathbf C_{l,s}\mathbf Z_{s,t-1}
   \end{split}
\end{equation}
\endgroup
where $\Y_t=(\Y_{1,t}^{\top},\ldots,\Y_{S,t}^{\top})^{\top}\in\mathbb R^N$ and $N=\sum_{s=1}^S N_s$.
The matrices $\mathbf W_{t-1}\in\mathbb R^{r_g\times N}$ and $\mathbf V_{s,t-1}\in\mathbb R^{r_l\times N_s}$ are the global and local portfolio weight matrices, respectively, while $\mathbf C_g\in\mathbb R^{r_g\times K}$ and $\mathbf C_{l,s}\in\mathbb R^{r_l\times K}$ are fixed characteristic-coefficient matrices. 
For group $s$, $\mathbf Z_{s,t-1}\in\mathbb R^{K\times N_s}$ contains transformed characteristics available before the period-$t$ returns are realized, and $\mathbf Z_{t-1}=(\mathbf Z_{1,t-1},\ldots,\mathbf Z_{S,t-1})\in\mathbb R^{K\times N}$ collects these characteristics across all groups.
The portfolio weights vary over time because $\mathbf Z_{t-1}$ and $\mathbf Z_{s,t-1}$ vary over time, even though $\mathbf C_g$ and $\mathbf C_{l,s}$ are fixed. This specification restricts each factor's portfolio weights to the span of the observed characteristics.

By substituting \eqref{eq:observable_factor_construction_intro} into the group factor model (\ref{eq:global_model_s_intro}), we propose a Hierarchical Group Factor Model (HGFM):
\begingroup\setlength{\abovedisplayskip}{4pt}\setlength{\belowdisplayskip}{4pt}\setlength{\abovedisplayshortskip}{2pt}\setlength{\belowdisplayshortskip}{2pt}
\begin{equation} \label{eq:proposed_model1_intro}
 \Y_{s,t}=\B_s\underbrace{(\mathbf C_g\mathbf Z_{t-1} \Y_t)}_{\text{Global: } \mathbf f_{g,t}}
 +\U_s\underbrace{(\mathbf C_{l,s}\mathbf Z_{s,t-1} \Y_{s,t})}_{\text{Local: }\mathbf f_{l,s,t}}
 +\boldsymbol\varepsilon_{s,t}.
\end{equation}
\endgroup

The global factor in  (\ref{eq:proposed_model1_intro}) is a portfolio of all assets with weights $\mathbf C_g\mathbf Z_{t-1}$, while the local factor for group $s$ uses only assets in that group, with weights $\mathbf C_{l,s}\mathbf Z_{s,t-1}$.
This linear construction can also represent sorting-based factors (e.g., the Fama--French size and value factors) when the rows of the characteristic matrices encode the stock weights of the constituent portfolios and the coefficient matrices specify their fixed long--short combinations. In our specification, the coefficient matrices are estimated jointly with the other HGFM parameters, allowing the data to determine how predetermined characteristics enter the global and local portfolio weights.

We first separate the global and local coefficient matrices conditional on the interaction matrices, normalize their factor representations, and impose a local information condition for statistical identification. Replacing latent factors with functions of observed returns introduces the outcome $\mathbf Y_t$ on the right-hand side of model~\eqref{eq:proposed_model1_intro}, which gives rise to endogeneity and generally renders ordinary least squares (OLS) estimators inconsistent. We therefore propose a quasi-maximum likelihood estimator (QMLE) computed by gradient ascent with an Armijo line search.

The asymptotic analysis allows both the cross-sectional dimension, $N$, and the time dimension, $T$, to diverge. Although the coefficient matrices are low-rank, the number of nuisance parameters increases with $N$, complicating the analysis of their asymptotic properties \citep{fan2004nonconcave}. We establish local consistency for the coefficient estimators and consistency and fixed-block asymptotic normality for the normalized factor parameters. The true ranks of the global and local factors are unknown, and standard eigenvalue-ratio methods \citep[e.g.,][]{lam2012factor,ahn2013eigenvalue} cannot be directly applied to select them in the present endogenous model. Following \citet*{ando2017clustering}, we propose an information criterion (IC) to jointly select the global rank and the common local rank, which consistently recovers the true ranks under suitable regularity assumptions.

Our model is related to \citet*{pu2025reduced}, who estimate reduced-rank spatio-temporal coefficient matrices by QMLE with an Armijo line search. Our approach differs in three respects. First, their specification uses an estimated weight matrix multiplying current and lagged returns, whereas ours constructs global and local factors from contemporaneous returns using characteristic-driven, time-varying weights. {Second, we jointly estimate the loading and characteristic-coefficient matrices for the global and local components, {allowing the characteristic coefficients of local factors to differ across groups while sharing the global factors.}} {Third, our hierarchical specification requires distinct identification conditions and a corresponding theoretical framework.} Conditional on the interaction matrices, cross-group characteristic variation separates the global and local coefficient matrices, and empirical characteristic second moments provide the basis for normalizing their factor representations. {We also develop an information criterion to jointly select the global rank and the common local rank.} Our asymptotic analysis treats the full characteristic design as fixed in advance and takes probability over the errors.

\paragraph*{Related literature.} The paper relates to three strands of research: group factor models, proxy-assisted factor estimation, and characteristic-based portfolio construction.

{The first strand concerns group factor models.} \citet{ando2017clustering} estimate group membership and group-specific latent factors in panels with observed predictors; our model takes group membership as given and estimates characteristic-driven factor weights. \citet*{hu2025aggregated} identify latent global factors by maximizing their average correlations with group factors, while \citet*{choi2023canonical} use cross-group canonical correlations to select latent global factors, with an application to U.S. industries. \citet{zhang2025multilevel} estimate global and local factors in matrix time series, while \citet{choi2023large} use global and national factors for large volatility matrix estimation. Related factor models also cover three-dimensional panels \citep{jin2025three}, matrix-valued time series \citep{wang2019factor}, and tensor observations \citep[e.g.,][]{chen2022factor,chen2024semi}. Factor structures of this kind also underlie high-dimensional covariance and portfolio estimation \citep{ding2021high}. The HGFM jointly estimates global and local coefficient matrices that map predetermined characteristics into time-varying portfolio weights, linking both factor components explicitly to observed characteristics and returns.

The second strand uses observable proxies to estimate latent factors. \citet{fan2021augmented} incorporate observed covariates into smoothed principal component analysis (PCA) to recover latent factors, while \citet*{wan2024mining} combine responses and a large set of factor proxies through penalized reduced-rank regression to estimate latent factors. \citet*{fan2022diversified} estimate latent factors using diversified cross-sectional projections, with individual characteristics among the proposed sources of portfolio weight matrices. Related studies examine common factor spaces \citep{andreou2025spanning}, the stability of observable--latent factor relations \citep{liao2024observable}, and observable-factor risk premia in the presence of omitted factors \citep{giglio2021asset}. The HGFM specifies both global and local factors as characteristic-weighted contemporaneous responses and jointly estimates the characteristic coefficients and loadings by QMLE to account for the resulting endogeneity.

The third strand concerns characteristic-based factor construction. \citet{kelly2019characteristics} use characteristics to parameterize time-varying factor loadings, and
{\citet*{feng2024deep} jointly learn nonlinear characteristic-based portfolio weights and factor loadings by minimizing aggregate realized pricing errors.}
\citet{brandt2009parametric} parameterize portfolio weights by firm characteristics and estimate their coefficients by maximizing average investor utility. 
{The HGFM specifies separate linear characteristic-based weighting rules for global and group-specific local factors.} {For this hierarchical specification, we provide identification conditions and QMLE inference for the weight coefficients and loadings.}

\paragraph*{Simulation and empirical highlights.} We examine the finite-sample accuracy of parameter estimation and rank selection through Monte Carlo simulations. We consider independent idiosyncratic errors under Gaussian, standardized-$t$, and mixture-normal distributions, as well as cross-sectional correlation and heteroskedasticity. Estimation errors generally decline with $T$, including under non-Gaussian errors, and rank-selection accuracy generally improves with $T$. Additional simulations examine robustness to sparse characteristic-driven weights, unequal local ranks, larger cross-sectional dimensions, and alternative penalty constants.

Empirically, we analyze daily returns on S\&P~100 constituents from five broad industry groups over 2016--2025. The analysis yields three main findings. First, the global factor assigns greater weight to larger stocks and stocks with higher book-to-market ratios, while {the characteristic coefficients of local factors differ across industries}. Second, combining global and local factors improves realized-return fit across all industry groups relative to either subset alone and the benchmark models. The combined model also explains average returns better than the benchmarks for the full stock panel, although its advantage varies across industries. Third, the combined portfolio achieves the highest annualized Sharpe ratio and the lowest volatility and maximum drawdown among the portfolios considered. Its Sharpe ratio remains above those of the equal- and value-weighted sample portfolios after transaction costs. Additional results examine sensitivity to industry grouping.

The article is structured as follows. Section~\ref{sec2} introduces the HGFM, its identification conditions, and the QMLE. Section~\ref{sec:theory} develops the asymptotic theory and rank-selection procedure. Section~\ref{sec3} evaluates finite-sample performance by Monte Carlo simulation. Section~\ref{sec4} presents the empirical application, and Section~\ref{sec5} concludes. The supplementary material contains proofs and additional simulation and empirical results.

\section{The Model and Methodology} \label{sec2}

In this section, we formally introduce the HGFM. We begin by specifying the model structure, in which the global and local factors are constructed from observed returns using characteristic-driven weights. We then discuss the identification of the global and local components and their factor representations, and detail the QMLE procedure. 

\subsection{Hierarchical Group Factor Model (HGFM)} 

By introducing global and local factors constructed from observed returns with characteristic-driven weights, we consider the following hierarchical group factor model:
\begingroup\setlength{\abovedisplayskip}{4pt}\setlength{\belowdisplayskip}{4pt}\setlength{\abovedisplayshortskip}{2pt}\setlength{\belowdisplayshortskip}{2pt}
\begin{equation} \label{eq:global_2}
	\begin{split}
		\begin{bmatrix}
			\mathbf Y_{1,t}\\
			\mathbf Y_{2,t}\\
			\vdots\\
			\mathbf Y_{S,t}
		\end{bmatrix}
		={}&
		\begin{bmatrix}
			\mathbf B_1\\
			\mathbf B_2\\
			\vdots\\
			\mathbf B_S
		\end{bmatrix}
		\mathbf C_g\mathbf Z_{t-1}\mathbf Y_t
		\\
		&+
		\begin{bmatrix}
			\mathbf U_1\mathbf C_{l,1}\mathbf Z_{1,t-1} & 0 & \cdots & 0\\
			0 & \mathbf U_2\mathbf C_{l,2}\mathbf Z_{2,t-1} & \cdots & 0\\
			\vdots & \vdots & \ddots & \vdots\\
			0 & 0 & \cdots & \mathbf U_S\mathbf C_{l,S}\mathbf Z_{S,t-1}
		\end{bmatrix}
		\begin{bmatrix}
			\mathbf Y_{1,t}\\
			\mathbf Y_{2,t}\\
			\vdots\\
			\mathbf Y_{S,t}
		\end{bmatrix}
		+
		\begin{bmatrix}
			\boldsymbol\varepsilon_{1,t}\\
			\boldsymbol\varepsilon_{2,t}\\
			\vdots\\
			\boldsymbol\varepsilon_{S,t}
		\end{bmatrix}.
	\end{split}
\end{equation}
\endgroup

In matrix form, Eq.~\eqref{eq:global_2} can be rewritten as
\begingroup\setlength{\abovedisplayskip}{4pt}\setlength{\belowdisplayskip}{4pt}\setlength{\abovedisplayshortskip}{2pt}\setlength{\belowdisplayshortskip}{2pt}
\begin{equation}\label{eq:global_2_compact}
	\begin{aligned}
		\mathbf Y_t
		&=
		\mathbf B\mathbf C_g\mathbf Z_{t-1}\mathbf Y_t
		+
		\operatorname{bdiag}\left(
		\mathbf U_1\mathbf C_{l,1}\mathbf Z_{1,t-1},
		\ldots,
		\mathbf U_S\mathbf C_{l,S}\mathbf Z_{S,t-1}
		\right)\mathbf Y_t
		+
		\boldsymbol\varepsilon_t
		\\
		&=
		\mathbf B\mathbf C_g\mathbf Z_{t-1}\mathbf Y_t
		+
		\mathbf U\mathbf C_l
		\operatorname{bdiag}\left(
		\mathbf Z_{1,t-1},
		\ldots,
		\mathbf Z_{S,t-1}
		\right)\mathbf Y_t
		+
		\boldsymbol\varepsilon_t
		\\
		&=
		\left\{
		\mathbf A_g\mathbf Z_{t-1}
		+
		\mathbf A_l
		\operatorname{bdiag}\left(
		\mathbf Z_{1,t-1},
		\ldots,
		\mathbf Z_{S,t-1}
		\right)
		\right\}\mathbf Y_t
		+
		\boldsymbol\varepsilon_t,
	\end{aligned}
\end{equation}
\endgroup
where $\mathbf B=(\mathbf B_1^\top,\ldots,\mathbf B_S^\top)^\top \in\mathbb R^{N\times r_g}$ denotes the global factor loading matrix, and $\mathbf U=\operatorname{bdiag}(\mathbf U_1,\ldots,\mathbf U_S)$ $\in\mathbb R^{N\times Sr_l}$ denotes the local factor loading matrix. 
Here, $\operatorname{bdiag}(\cdot)$ forms a block-diagonal matrix from its arguments.
Moreover, $\mathbf C_l=\operatorname{bdiag}(\mathbf C_{l,1},\ldots,\mathbf C_{l,S})$
	$\in\mathbb R^{Sr_l\times SK}$ is the block-diagonal matrix of local characteristic coefficients, 
	and $\boldsymbol\varepsilon_t=(\boldsymbol\varepsilon_{1,t}^\top,\ldots,
	\boldsymbol\varepsilon_{S,t}^\top)^\top$ is the idiosyncratic error vector.
	Define $\mathbf A_g=\mathbf B\mathbf C_g\in\mathbb R^{N\times K}$,
	$\mathbf A_{l,s}=\mathbf U_s\mathbf C_{l,s}\in\mathbb R^{N_s\times K}$,
	and $\mathbf A_l=\mathbf U\mathbf C_l
	=\operatorname{bdiag}(\mathbf A_{l,1},\ldots,\mathbf A_{l,S})
	\in\mathbb R^{N\times SK}$.
{
Define the time-varying coefficient matrix $\mathbf M_t=\mathbf A_g\mathbf Z_{t-1}
+\operatorname{bdiag}\{\mathbf A_{l,s}\mathbf Z_{s,t-1}\}_{s=1}^S$.
Then model~\eqref{eq:global_2_compact} is $\mathbf Y_t=\mathbf M_t\mathbf Y_t+\boldsymbol\varepsilon_t$. We refer to $\mathbf M_t$ as the interaction matrix because it describes contemporaneous dependence among the observed units.
}
When $K$, $r_g$, $r_l$, and $S$ are fixed, the number of unknown parameters associated with $\mathbf A_g$ and $\mathbf A_l$ grows linearly with $N$. This linear growth follows from the fixed dimension of the characteristics, while the low-rank restrictions further constrain the global and local coefficient matrices. Under suitable conditions, the model can be consistently estimated as $T$ diverges.

\begin{remark}[{Relation to spatial models}]
Our model \eqref{eq:global_2_compact} is related to spatial autoregressive (SAR) models, as both capture contemporaneous dependence. Traditional SAR models rely on a prespecified spatial weight matrix, whereas the HGFM uses predetermined characteristics and estimated low-rank coefficient matrices to construct time-varying interaction matrices.
The model decomposes these interactions into global and local components. Setting $\B=0$ leaves only within-group interactions, while setting $\U=0$ yields a purely global structure.
The model therefore represents dependence both across and within groups through observed characteristics.
\end{remark}

\subsection{Model Identifiability} 

Identification issues commonly arise in latent factor models \citep{bai2003inferential,bai2012statistical,bai2013principal,Wang2022GFM}.
The proposed model raises two identification issues: the separation of the global and group-specific components and the factorization of each low-rank coefficient matrix. We address these two issues sequentially. First, we exploit cross-group characteristic variation to separate the global and local coefficient matrices. We then impose a Singular Value Decomposition (SVD) normalization on the low-rank factors of each separated coefficient matrix.

The first identification issue concerns whether the time-invariant global and local coefficient matrices can be separately recovered from the time-varying interaction matrix. To illustrate this issue, let
 $\mathbf x_{s,t}=\mathbf Z_{s,t-1}\mathbf Y_{s,t}$ and
	$\mathbf x_{-s,t}=\sum_{j\ne s}\mathbf Z_{j,t-1}\mathbf Y_{j,t}$, and let
	$\mathbf A_{g,s}$ denote the row block of $\mathbf A_g$ corresponding to
	group $s$. The model for group $s$ can then be written as
	${\mathbf Y_{s,t}=\mathbf A_{g,s}\mathbf x_{-s,t}
		+(\mathbf A_{g,s}+\mathbf A_{l,s})\mathbf x_{s,t}
		+\boldsymbol\varepsilon_{s,t}.}$
	Thus, the cross-group input $\mathbf x_{-s,t}$ is governed exclusively by
	$\mathbf A_{g,s}$, whereas the own-group input $\mathbf x_{s,t}$ is governed
	by $\mathbf A_{g,s}+\mathbf A_{l,s}$. 
Given identification of the time-varying interaction matrix, this structure allows us to distinguish $\mathbf A_{g,s}$ and $\mathbf A_{l,s}$ when the characteristics vary sufficiently both within and outside group $s$.

Unless otherwise stated, we treat the entire characteristic array as a
nonstochastic design fixed in advance throughout the theoretical analysis.
For each group $s$, let $\mathbf Z_{-s,t-1} \in\mathbb R^{K\times(N-N_s)}$ denote the
	concatenation of $\{\mathbf Z_{j,t-1}:j\ne s\}$, and define the corresponding design matrices by
	\begingroup\setlength{\abovedisplayskip}{4pt}\setlength{\belowdisplayskip}{4pt}\setlength{\abovedisplayshortskip}{2pt}\setlength{\belowdisplayshortskip}{2pt}
\[
	\mathbf Q_{Z,s,T}
	=
	\frac{1}{TN_s}\sum_{t=1}^T
	\mathbf Z_{s,t-1}\mathbf Z_{s,t-1}^{\top},
	\qquad
	\mathbf Q_{Z,-s,T}
	=
	\frac{1}{T(N-N_s)}\sum_{t=1}^T
	\mathbf Z_{-s,t-1}\mathbf Z_{-s,t-1}^{\top}.
	\]
\endgroup
	The following condition ensures that the characteristic variation is
	{nondegenerate both within group $s$ and across the groups outside $s$.}
    
	\begin{condition}
		\label{ass:feature_variation}
		There exist constants $0<c_Z<C_Z<\infty$ such that, for every
		$s=1,\ldots,S$,
		\begingroup\setlength{\abovedisplayskip}{4pt}\setlength{\belowdisplayskip}{4pt}\setlength{\abovedisplayshortskip}{2pt}\setlength{\belowdisplayshortskip}{2pt}
\[
		\begin{aligned}
			c_Z
			&\leq \lambda_{\min}(\mathbf Q_{Z,s,T})
			\leq \lambda_{\max}(\mathbf Q_{Z,s,T})
			\leq C_Z,\\
			c_Z
			&\leq \lambda_{\min}(\mathbf Q_{Z,-s,T})
			\leq \lambda_{\max}(\mathbf Q_{Z,-s,T})
			\leq C_Z.
		\end{aligned}
		\]
\endgroup
	\end{condition}
	
Condition~\ref{ass:feature_variation} rules out degenerate characteristic directions that would prevent the global and local components from being distinguished. 
The following proposition states that if two collections generate
the same interaction matrices at the design dates, their global and
group-specific coefficient matrices coincide.

	\begin{proposition}
		\label{prop:feature_global_local_separation}
		Suppose that Condition~\ref{ass:feature_variation} holds. Consider two
		collections of coefficient matrices
		$\{\mathbf A_g,\mathbf A_{l,s}:s=1,\ldots,S\}$ and
		$\{\widetilde{\mathbf A}_g,\widetilde{\mathbf A}_{l,s}:s=1,\ldots,S\}$.
		For every $t=1,\ldots,T$, if
		\begingroup\setlength{\abovedisplayskip}{4pt}\setlength{\belowdisplayskip}{4pt}\setlength{\abovedisplayshortskip}{2pt}\setlength{\belowdisplayshortskip}{2pt}
\[
		\mathbf A_g\mathbf Z_{t-1}
		+
		\operatorname{bdiag}\!\left\{
		\mathbf A_{l,s}\mathbf Z_{s,t-1}
		\right\}_{s=1}^S
		=
		\widetilde{\mathbf A}_g\mathbf Z_{t-1}
		+
		\operatorname{bdiag}\!\left\{
		\widetilde{\mathbf A}_{l,s}\mathbf Z_{s,t-1}
		\right\}_{s=1}^S,
		\]
\endgroup
		then $\mathbf A_g=\widetilde{\mathbf A}_g$ and
		$\mathbf A_{l,s}=\widetilde{\mathbf A}_{l,s}$ for every
		$s=1,\ldots,S$.
	\end{proposition}
	
Under Condition~\ref{ass:feature_variation} and conditional on the interaction matrices, Proposition~\ref{prop:feature_global_local_separation} separates $\mathbf A_{g,s}$ through cross-group characteristic variation and then $\mathbf A_{l,s}$ through within-group variation, for every $s=1,\ldots,S$.

The second identification issue is the usual rotation indeterminacy of low-rank factor models
	\citep{bai2012statistical,ouyang2026statistical}. For any nonsingular
	matrices $\mathbf H_g$ and $\mathbf H_{l,s}$, the transformations
	$\widetilde{\mathbf B}=\mathbf B\mathbf H_g$,
	$\widetilde{\mathbf C}_g=\mathbf H_g^{-1}\mathbf C_g$,
	$\widetilde{\mathbf U}_s=\mathbf U_s\mathbf H_{l,s}$, and
	$\widetilde{\mathbf C}_{l,s}
	=\mathbf H_{l,s}^{-1}\mathbf C_{l,s}$
	leave $\mathbf B\mathbf C_g$ and
	$\mathbf U_s\mathbf C_{l,s}$ unchanged. Thus, the factor representations
	remain unidentified even after the global and local coefficient matrices
	have been separated. We therefore impose an SVD normalization to
	identify the factors without changing the corresponding coefficient
	matrices.
    
Define the empirical second-moment matrices of the observed characteristics as
\begingroup\setlength{\abovedisplayskip}{4pt}\setlength{\belowdisplayskip}{4pt}\setlength{\abovedisplayshortskip}{2pt}\setlength{\belowdisplayshortskip}{2pt}
\[
\boldsymbol\Gamma_{Z,T}
=
\frac{1}{T}\sum_{t=1}^T
\mathbf Z_{t-1}\mathbf Z_{t-1}^{\top},
\qquad
\boldsymbol\Gamma_{Z,s,T}
=
\frac{1}{T}\sum_{t=1}^T
\mathbf Z_{s,t-1}\mathbf Z_{s,t-1}^{\top}.
\]
\endgroup
Since $\boldsymbol\Gamma_{Z,T}=\sum_{s=1}^S N_s\mathbf Q_{Z,s,T}$
and $\boldsymbol\Gamma_{Z,s,T}=N_s\mathbf Q_{Z,s,T}$,
both matrices are positive definite under Condition~\ref{ass:feature_variation}. Using these matrices, the following normalization condition identifies each low-rank factorization up to paired sign changes.
	
	\begin{condition}
		\label{cond:internal_factor_identification}
		The global and local factor matrices satisfy
		$\mathbf B^{\top}\mathbf B
		=\mathbf C_g\boldsymbol\Gamma_{Z,T}\mathbf C_g^{\top}
		=\mathbf D_g$ and
		$\mathbf U_s^{\top}\mathbf U_s
		=\mathbf C_{l,s}\boldsymbol\Gamma_{Z,s,T}
		\mathbf C_{l,s}^{\top}
		=\mathbf D_{l,s}$ for every $s=1,\ldots,S$, where
		$\mathbf D_g$ and $\mathbf D_{l,s}$ are diagonal matrices whose diagonal
		entries are positive, distinct, and arranged in decreasing order.
	\end{condition}
	
Condition~\ref{cond:internal_factor_identification} adopts the standard balanced normalization used in factor models \citep{bai2012statistical,bai2013principal, Wang2022GFM,ouyang2026statistical}.
The balancing equality fixes relative scale, while diagonality, distinctness, and decreasing ordering remove rotation and permutation indeterminacies, leaving only paired sign changes.

	\begin{proposition}
		\label{prop:internal_factor_identification}
		Suppose that
		$\mathbf A_g=\mathbf B\mathbf C_g
		=\widetilde{\mathbf B}\widetilde{\mathbf C}_g$ and
		$\mathbf A_{l,s}=\mathbf U_s\mathbf C_{l,s}
		=\widetilde{\mathbf U}_s\widetilde{\mathbf C}_{l,s}$ for every
		$s=1,\ldots,S$. If both representations satisfy
		Condition~\ref{cond:internal_factor_identification}, then there exist
		diagonal sign matrices $\mathbf S_g$ and $\mathbf S_{l,s}$, with diagonal
		entries in $\{-1,1\}$, such that
		\begingroup\setlength{\abovedisplayskip}{4pt}\setlength{\belowdisplayskip}{4pt}\setlength{\abovedisplayshortskip}{2pt}\setlength{\belowdisplayshortskip}{2pt}
\[
		\widetilde{\mathbf B}=\mathbf B\mathbf S_g,
		\quad
		\widetilde{\mathbf C}_g=\mathbf S_g\mathbf C_g,
		\quad
		\widetilde{\mathbf U}_s=\mathbf U_s\mathbf S_{l,s},
		\quad
		\widetilde{\mathbf C}_{l,s}
		=\mathbf S_{l,s}\mathbf C_{l,s}.
		\]
\endgroup
	\end{proposition}
	
Proposition~\ref{prop:internal_factor_identification} establishes the uniqueness of the normalized factorization up to paired sign changes. 
To complete the factorization argument for coefficient matrices
with the specified exact ranks and singular-value gaps, we construct a
normalized representation by an ordered singular value decomposition.

	\begin{proposition}
		\label{prop:svd_internal_transformation}
		Suppose that Condition~\ref{ass:feature_variation} holds. Assume that $\operatorname{rank}(\mathbf A_g)=r_g$ and
$\operatorname{rank}(\mathbf A_{l,s})=r_l$ for every $s$. Assume that the
		positive singular values of
		$\mathbf A_g\boldsymbol\Gamma_{Z,T}^{1/2}$ are distinct and that, for every
		$s=1,\ldots,S$, those of
		$\mathbf A_{l,s}\boldsymbol\Gamma_{Z,s,T}^{1/2}$ are also distinct. Let
		  $\mathbf A_g\boldsymbol\Gamma_{Z,T}^{1/2}
		=\mathbf P_g\boldsymbol\Lambda_g\mathbf V_g^{\top}$
		 be its ordered thin singular value decomposition, and define
		$\mathbf B^{\dagger}=\mathbf P_g\boldsymbol\Lambda_g^{1/2}$ and
		  $\mathbf C_g^{\dagger}
		=\boldsymbol\Lambda_g^{1/2}\mathbf V_g^{\top}
		\boldsymbol\Gamma_{Z,T}^{-1/2}$.
		For each $s$, let
		  $\mathbf A_{l,s}\boldsymbol\Gamma_{Z,s,T}^{1/2}
		=\mathbf P_{l,s}\boldsymbol\Lambda_{l,s}\mathbf V_{l,s}^{\top}$
		 be its ordered thin singular value decomposition, and define
		$\mathbf U_s^{\dagger}
		=\mathbf P_{l,s}\boldsymbol\Lambda_{l,s}^{1/2}$ and
		  $\mathbf C_{l,s}^{\dagger}
		=\boldsymbol\Lambda_{l,s}^{1/2}\mathbf V_{l,s}^{\top}
		\boldsymbol\Gamma_{Z,s,T}^{-1/2}$.
		Then $\mathbf B^{\dagger}\mathbf C_g^{\dagger}=\mathbf A_g$ and
		$\mathbf U_s^{\dagger}\mathbf C_{l,s}^{\dagger}
		=\mathbf A_{l,s}$ for every $s=1,\ldots,S$. Moreover, the constructed
		factors satisfy Condition~\ref{cond:internal_factor_identification} with
		$\mathbf D_g=\boldsymbol\Lambda_g$ and
		$\mathbf D_{l,s}=\boldsymbol\Lambda_{l,s}$.
	\end{proposition}

	For coefficient matrices of the stated exact ranks and with
distinct positive singular values,
Proposition~\ref{prop:svd_internal_transformation} constructs the
normalized factorization and
Proposition~\ref{prop:internal_factor_identification} gives uniqueness
up to paired signs. 
Together with Proposition~\ref{prop:feature_global_local_separation}, these results establish the uniqueness of the coefficient decomposition and its normalized factorization for the given interaction matrices.

\begin{remark}[{Equivariance to the ordering of the observed units}]  
	If assets are reordered within the known groups, the responses and columns of the characteristic matrices must be reordered jointly. If the groups are reordered, the group partition must be permuted as well. {The rows of the loading matrices follow the same reordering as the assets. The characteristic coefficients and factor series are preserved when the same sign convention is used. This is because jointly reordering returns and characteristics leaves their product unchanged. The normalization therefore does not depend on which assets appear first and does not require anchor assets.}
\end{remark}

\subsection{Quasi-maximum Likelihood Estimator}  

Because $\mathbf Y_t$ appears on both sides of
model~\eqref{eq:global_2_compact}, the contemporaneous regressors depend on the errors, so ordinary least squares is generally inconsistent.
The ranks $r_g$ and $r_l$ are treated as known throughout this subsection.
We therefore adopt the quasi-maximum likelihood approach of
\citet{lee2004asymptotic}.
Let $\boldsymbol\theta=(\mathbf B,\mathbf C_g,
	\{\mathbf U_s,\mathbf C_{l,s}\}_{s=1}^S)$ collect the factor
parameters before normalization, and let $\boldsymbol\tau=(\tau_1,\ldots,\tau_N)^\top$, $\tau_i=\log\sigma_i^2$, and $\boldsymbol\Phi(\boldsymbol\tau)
	=\operatorname{diag}(e^{\tau_1},\ldots,e^{\tau_N})$.
The associated coefficient parameter is $\boldsymbol\eta(\boldsymbol\theta,\boldsymbol\tau)
	=(\mathbf B\mathbf C_g,
	\{\mathbf U_s\mathbf C_{l,s}\}_{s=1}^S,
	\boldsymbol\tau)$.
	Let
	$\mathbf M_t(\boldsymbol\theta)$ be the interaction matrix in
	model~\eqref{eq:global_2_compact} generated by
	$\mathbf B\mathbf C_g$ and
	$\{\mathbf U_s\mathbf C_{l,s}\}_{s=1}^S$, and define
	$\mathbf S_t(\boldsymbol\theta)
	=\mathbf I_N-\mathbf M_t(\boldsymbol\theta)$.
	Up to an additive constant independent of the parameters, the sample-size-normalized negative  quasi-log-likelihood is
	\begingroup\setlength{\abovedisplayskip}{4pt}\setlength{\belowdisplayskip}{4pt}\setlength{\abovedisplayshortskip}{2pt}\setlength{\belowdisplayshortskip}{2pt}
\begin{equation}
		\label{eq:factor_negative_qll_v69}
		\begin{aligned}
			Q_{T,F}(\boldsymbol\theta,\boldsymbol\tau)
			={}&
			\frac{1}{2N}\sum_{i=1}^N\tau_i
			-\frac{1}{NT}\sum_{t=1}^T
			\log\left|\mathbf S_t(\boldsymbol\theta)\right|\\
			&+
			\frac{1}{2NT}\sum_{t=1}^T
			\mathbf Y_t^\top
			\mathbf S_t(\boldsymbol\theta)^\top
			\boldsymbol\Phi(\boldsymbol\tau)^{-1}
			\mathbf S_t(\boldsymbol\theta)
			\mathbf Y_t.
		\end{aligned}
	\end{equation}
\endgroup

Let $\mathcal V_N(r_g,r_l)$ denote the set of $(\boldsymbol\theta,\boldsymbol\tau)$ whose coefficient products and variance vector satisfy the compactness, rank, and stability restrictions in Condition~\ref{ass:parameter_space_stability}.
Estimation proceeds by minimizing $Q_{T,F}$ over this parameter space.

The variance parameters can be concentrated out of the criterion.
For fixed $\boldsymbol\theta$, the variance first-order conditions  give
	\begingroup\setlength{\abovedisplayskip}{4pt}\setlength{\belowdisplayskip}{4pt}\setlength{\abovedisplayshortskip}{2pt}\setlength{\belowdisplayshortskip}{2pt}
\begin{equation}
	\label{eq:sigma}
	\widehat\sigma_i^2(\boldsymbol\theta)
	=\frac{1}T\sum_{t=1}^T
	\{(\boldsymbol\iota_i^{(N)})^\top
	\mathbf S_t(\boldsymbol\theta)\mathbf Y_t\}^2,
	\qquad i=1,\ldots,N,
	\end{equation}
\endgroup 
	  where $\boldsymbol\iota_j^{(d)}$ denotes the $j$th canonical vector in
	$\mathbb R^d$. Substituting \eqref{eq:sigma} into
	\eqref{eq:factor_negative_qll_v69}, multiplying by \(-NT\),   and omitting the resulting constant gives the concentrated quasi-log-likelihood
	\begingroup\setlength{\abovedisplayskip}{4pt}\setlength{\belowdisplayskip}{4pt}\setlength{\abovedisplayshortskip}{2pt}\setlength{\belowdisplayshortskip}{2pt}
\begin{equation}
		\mathcal L_{T,c}(\boldsymbol\theta)
		=
		\sum_{t=1}^T
		\log\left|\mathbf S_t(\boldsymbol\theta)\right|
		-\frac{T}{2}\sum_{i=1}^N
		\log\widehat\sigma_i^2(\boldsymbol\theta).
	\end{equation}
\endgroup

Following \citet{he2021} and \citet{pu2025reduced}, we maximize
	$\mathcal L_{T,c}$ over the factor matrices using a gradient-based
	algorithm with an Armijo line search. Denote the retained, not-yet-normalized
	factor matrices by $\widetilde{\boldsymbol\theta}$. Neither singular-value truncation nor the SVD normalization is imposed
	during the iterations. In
	implementation, the optimization is repeated from multiple full-rank
	initial values, and the candidate attaining the largest value of
	$\mathcal L_{T,c}$ is retained.
	The corresponding log-variance estimates are
	$\widehat\tau_i
	=\log\widehat\sigma_i^2
	(\widetilde{\boldsymbol\theta})$.

After optimization, define $\widehat{\mathbf A}_g
	=\widetilde{\mathbf B}\widetilde{\mathbf C}_g$
	and
	$\widehat{\mathbf A}_{l,s}
	=\widetilde{\mathbf U}_s\widetilde{\mathbf C}_{l,s}$.
We then apply the post-estimation normalization in Proposition~\ref{prop:svd_internal_transformation} to these estimated coefficient matrices, using $\boldsymbol\Gamma_{Z,T}$ and $\boldsymbol\Gamma_{Z,s,T}$ for the global and local components, respectively, obtaining $\widehat{\mathbf B}$, $\widehat{\mathbf C}_g$, $\widehat{\mathbf U}_s$, and $\widehat{\mathbf C}_{l,s}$.
By construction, $\widehat{\mathbf B}\widehat{\mathbf C}_g =\widehat{\mathbf A}_g$ and $\widehat{\mathbf U}_s\widehat{\mathbf C}_{l,s} =\widehat{\mathbf A}_{l,s}$ for every $s=1,\ldots,S$.
Consequently, the normalization leaves the estimated coefficient matrices, variance estimates, and quasi-likelihood value unchanged while selecting an identified factor representation.

\IncMargin{1em}
\begin{algorithm}[h!]
	\linespread{0.8}\selectfont
	\caption{Gradient algorithm with Armijo line search\label{al1}}
	\SetAlgoNoLine
	\SetKwInOut{Input}{\textbf{Input}}
	\SetKwInOut{Output}{\textbf{Output}}
	\Input{Group returns \(\{\mathbf Y_{s,t}\}\); predetermined characteristics
		\(\{\mathbf Z_{s,t-1}\}\); ranks \(r_g,r_l\); $M$ initial values
		\(\{\boldsymbol\theta^{(0,m)}\}_{m=1}^M\), where
		\(\boldsymbol\theta=(\mathbf B,\mathbf C_g,
		\{\mathbf U_s,\mathbf C_{l,s}\}_{s=1}^S)\); maximum numbers of iterations
		\(K_{\mathrm{out}},K_{\mathrm{in}}\); scale factor \(\varpi\);
		tolerance \(\chi\); Armijo parameter \(\xi\).\\}
	\Output{\(\widehat{\boldsymbol\sigma}^{\,2},\widehat{\mathbf B},
		\widehat{\mathbf C}_g,
		\{\widehat{\mathbf U}_s,\widehat{\mathbf C}_{l,s}\}_{s=1}^S\).\\}
	\BlankLine
	\For{\(m=1,\ldots,M\)}{
		\For{\(k=0,\ldots,K_{\mathrm{out}}-1\)}{
			Compute \(\widehat{\boldsymbol\sigma}^{\,2(k,m)}\) from
			\eqref{eq:sigma}, \(\mathcal L_{T,c}(\boldsymbol\theta^{(k,m)})\),
			and \(\mathbf g^{(k,m)}=\nabla_{\boldsymbol\theta}
			\mathcal L_{T,c}(\boldsymbol\theta^{(k,m)})\)\;
			\If{\(\|\mathbf g^{(k,m)}\|_\infty<\chi\)}{
				Mark the current estimate as converged and \textbf{break}\;
			}
			\For{\(i=1,\ldots,K_{\mathrm{in}}\)}{
				Set \(a_i=\varpi^{i-1}\) and
				\(\boldsymbol\theta_{i}^{(k+1,m)}=
				\boldsymbol\theta^{(k,m)}+a_i\mathbf g^{(k,m)}\)\;
				\If{\(\mathcal L_{T,c}(\boldsymbol\theta_i^{(k+1,m)})\geq
					\mathcal L_{T,c}(\boldsymbol\theta^{(k,m)})+
					\xi a_i\|\mathbf g^{(k,m)}\|_2^2\)}{
					Set \(\boldsymbol\theta^{(k+1,m)}=
					\boldsymbol\theta_i^{(k+1,m)}\) and \textbf{break}\;
				}
			}
		}
	}
   Among the admissible converged terminal estimates, retain the one with the largest
	  profile likelihood; denote its factors by \(\widetilde{\boldsymbol\theta}\) \;
	Apply the normalization in Proposition~\ref{prop:svd_internal_transformation} to
	 the retained coefficient matrices
	  \(\widetilde{\mathbf B}\widetilde{\mathbf C}_g\) and
	\(\widetilde{\mathbf U}_s\widetilde{\mathbf C}_{l,s}\),
	using $\boldsymbol\Gamma_{Z,T}$ and $\boldsymbol\Gamma_{Z,s,T}$, respectively,
	and denote the
	normalized factors by
\(\widehat{\mathbf B},\widehat{\mathbf C}_g,
\{\widehat{\mathbf U}_s,\widehat{\mathbf C}_{l,s}\}_{s=1}^S\) \;
	{\textbf{return}} \(\widehat{\boldsymbol\sigma}^{\,2},\widehat{\mathbf B},
	\widehat{\mathbf C}_g,
	\{\widehat{\mathbf U}_s,\widehat{\mathbf C}_{l,s}\}_{s=1}^S\)\;
\end{algorithm}
\DecMargin{1em}

\begin{remark}[Scope of the numerical guarantee]  Algorithm~\ref{al1} produces numerical candidate estimates from $M$ initial values. The Armijo line search increases the objective along accepted updates, but neither the line search nor multiple starts guarantees a global optimum.  The theory below establishes the existence and rate of a local solution in a neighborhood of the identified parameter. 
\end{remark}

\section{Theoretical Results}\label{sec:theory} 
Under the stated conditions, we establish local consistency for the coefficient and log-variance estimators, transfer this rate to the normalized factor representation, and derive asymptotic normality for prespecified loading rows and characteristic-coefficient columns. We then establish consistency of joint global and common local rank selection under suitable penalty rates.

\subsection{Average Consistency}
	
Before establishing the average consistency of the proposed QMLE, we introduce some notation. For a generic matrix $\mathbf A=(a_{ij})\in\mathbb R^{m\times n}$, $\operatorname{vec}(\mathbf A)$ denotes its column-wise vectorization. 
Let $\lambda_{\min} (\textbf{A})$ and $\lambda_{\max} (\textbf{A})$ denote the minimum and maximum eigenvalues of a symmetric matrix $\textbf{A}$, respectively. In addition, $\|\textbf{A}\|_q$ denotes the induced matrix $q$-norm for $q=1,2,\infty$, and $\|\textbf{A}\|_F=(\sum_{i=1}^m\sum_{j=1}^n a_{ij}^2)^{1/2}$ denotes the Frobenius norm.
	
	Let $\boldsymbol\eta=(\mathbf A_g,
	\{\mathbf A_{l,s}\}_{s=1}^S,\boldsymbol\tau)$ denote the coefficient and
	log-variance parameters, and let $\boldsymbol\eta_v$ denote the column-wise
	vectorization of
	$(\mathbf A_g,\mathbf A_{l,1},\ldots,\mathbf A_{l,S},
	\boldsymbol\tau)$.
	The corresponding scaling matrix is
	$\mathbf D_{\eta,N}=\operatorname{bdiag}\{\mathbf I_{NK},
	\{\mathbf I_{N_sK}\}_{s=1}^S,N\mathbf I_N\}$.
	For a vector $\mathbf v$ with the same dimension as
	$\boldsymbol\eta_v$, set $\|\mathbf v\|_N^2
	=\mathbf v^\top\mathbf D_{\eta,N}^{-1}\mathbf v$.
	For two coefficient-space parameters $\boldsymbol\eta_a$ and
	$\boldsymbol\eta_b$,   use the same notation for their vectorized difference:
	\begingroup\setlength{\abovedisplayskip}{4pt}\setlength{\belowdisplayskip}{4pt}\setlength{\abovedisplayshortskip}{2pt}\setlength{\belowdisplayshortskip}{2pt}
\[
	\|\boldsymbol\eta_a-\boldsymbol\eta_b\|_N^2
	=
	\|\mathbf A_{g,a}-\mathbf A_{g,b}\|_F^2
	+
	\sum_{s=1}^S
	\|\mathbf A_{l,s,a}-\mathbf A_{l,s,b}\|_F^2
	+
	N^{-1}\|\boldsymbol\tau_a-\boldsymbol\tau_b\|_2^2.
	\]
\endgroup
	Equivalently,
	$\|\boldsymbol\eta_a-\boldsymbol\eta_b\|_N^2
	=(\boldsymbol\eta_{v,a}-\boldsymbol\eta_{v,b})^\top
	\mathbf D_{\eta,N}^{-1}
	(\boldsymbol\eta_{v,a}-\boldsymbol\eta_{v,b})$.
	
	For average consistency, we use the coefficient parameter
	$\boldsymbol\eta$. Define
	$\mathbf S_t(\boldsymbol\eta)
	=\mathbf I_N-\mathbf M_t(\boldsymbol\eta)$, where
	$\mathbf M_t(\boldsymbol\eta)$ is the interaction matrix in
	\eqref{eq:global_2_compact} written in terms of
	$\mathbf A_g$ and $\{\mathbf A_{l,s}\}_{s=1}^S$.
	Since $Q_{T,F}$ depends on the direct factors only through
	$\mathbf A_g=\mathbf B\mathbf C_g$ and
	$\mathbf A_{l,s}=\mathbf U_s\mathbf C_{l,s}$, its equivalent
	coefficient-space form is
	\begingroup\setlength{\abovedisplayskip}{4pt}\setlength{\belowdisplayskip}{4pt}\setlength{\abovedisplayshortskip}{2pt}\setlength{\belowdisplayshortskip}{2pt}
\begin{equation}
		\begin{aligned}
			Q_T(\boldsymbol\eta)
			={}&
			\frac{1}{2N}\sum_{i=1}^N\tau_i
			-\frac{1}{NT}\sum_{t=1}^T
			\log\left|\mathbf S_t(\boldsymbol\eta)\right|\\
			&+
			\frac{1}{2NT}\sum_{t=1}^T
			\mathbf Y_t^\top
			\mathbf S_t(\boldsymbol\eta)^\top
			\boldsymbol\Phi(\boldsymbol\tau)^{-1}
			\mathbf S_t(\boldsymbol\eta)
			\mathbf Y_t.
		\end{aligned}
	\end{equation}
\endgroup

The admissible coefficient space $\mathcal H_N(r_g,r_l)$ is the compact, rank-restricted, and stable set specified in Condition~\ref{ass:parameter_space_stability} below.
Define the population criterion by
$Q(\boldsymbol\eta)
=\mathrm E\{Q_T(\boldsymbol\eta)\}$ on the admissible coefficient space.
Let $\boldsymbol\eta^0$ denote the data-generating parameter; it minimizes
the population criterion.

We next state the regularity conditions.

\begin{condition}
\label{ass:parameter_space_stability}
Let $\mathcal H_N(r_g,r_l)$ be a compact parameter space for
$\boldsymbol\eta=(\mathbf A_g,\{\mathbf A_{l,s}\}_{s=1}^S,
\boldsymbol\tau)$, with
$\operatorname{rank}(\mathbf A_g)\leq r_g$ and
$\operatorname{rank}(\mathbf A_{l,s})\leq r_l$ for each $s$.
There exist constants $0<C_A,C_S<\infty$, 
such that, for all $\boldsymbol\eta\in\mathcal H_N(r_g,r_l)$,
\begingroup\setlength{\abovedisplayskip}{4pt}\setlength{\belowdisplayskip}{4pt}\setlength{\abovedisplayshortskip}{2pt}\setlength{\belowdisplayshortskip}{2pt}
\[
N\|\mathbf A_g\|_F^2
+\sum_{s=1}^S N_s\|\mathbf A_{l,s}\|_F^2
\leq C_A,
\qquad
\|\boldsymbol\tau\|_\infty\leq C_A.
\]
\endgroup
Moreover, $\mathbf S_t(\boldsymbol\eta)$ is nonsingular for every
$t=1,\ldots,T$, and
\begingroup\setlength{\abovedisplayskip}{4pt}\setlength{\belowdisplayskip}{4pt}\setlength{\abovedisplayshortskip}{2pt}\setlength{\belowdisplayshortskip}{2pt}
\[
\sup_{\boldsymbol\eta\in\mathcal H_N(r_g,r_l)}
\max_{1\leq t\leq T}
\left\{
\|\mathbf S_t(\boldsymbol\eta)\|_1
+\|\mathbf S_t(\boldsymbol\eta)\|_\infty
+\|\mathbf S_t(\boldsymbol\eta)^{-1}\|_1
+\|\mathbf S_t(\boldsymbol\eta)^{-1}\|_\infty
\right\}
\leq C_S.
\]
\endgroup
\end{condition}

	\begin{condition}
		\label{ass:true_parameter_interior}
The true coefficient matrices have ranks $r_g$ and $r_l$, and $\boldsymbol\eta^0$   is uniformly interior to the non-rank restrictions in Condition~\ref{ass:parameter_space_stability}.
At the normalized factor representation, let $\mathbf D_g^0=(\mathbf B^0)^\top\mathbf B^0
		=\mathbf C_g^0\boldsymbol\Gamma_{Z,T}(\mathbf C_g^0)^\top$ and write
		$\mathbf D_g^0=\operatorname{diag}(d_{g,1},\ldots,d_{g,r_g})$.
		For each $s$, let
		$\mathbf D_{l,s}^0=(\mathbf U_s^0)^\top\mathbf U_s^0
		=\mathbf C_{l,s}^0\boldsymbol\Gamma_{Z,s,T}
		(\mathbf C_{l,s}^0)^\top$ and write
		$\mathbf D_{l,s}^0
		=\operatorname{diag}(d_{l,s,1},\ldots,d_{l,s,r_l})$.
		There exist constants $0<c_D<C_D<\infty$ and
		$c_{\mathrm{gap}}>0$, independent of $N,T$, such that every diagonal
		entry of these matrices lies in $[c_D,C_D]$ and every adjacent difference
		within each matrix is at least $c_{\mathrm{gap}}$.
	\end{condition}

	\begin{condition}
		\label{ass:balanced_groups}
		The number of groups $S \geq 2$ is fixed, and there exist constants
		$0<c_{\mathrm{grp}} < C_{\mathrm{grp}}<\infty$ such that $c_{\mathrm{grp}}\leq N_s/N\leq C_{\mathrm{grp}}$ for every
		$s=1,\ldots,S$.
	\end{condition}
	
	\begin{condition}
		\label{ass:errors}
        For some constant $C_{\mathrm{char}}<\infty$, the characteristic array satisfies
$\sup_{N,T}\max_{1\leq t\leq T}
\max_{1\leq k\leq K,\,1\leq i\leq N}|Z_{ki,t-1}|
\leq C_{\mathrm{char}}$. The error vectors are independent over time,
and their entries are mutually independent within each date. For some
$\delta>0$ and every $t=1,\ldots,T$,
$\mathrm E(\boldsymbol\varepsilon_t)=\mathbf0$,
$\mathrm E(\boldsymbol\varepsilon_t\boldsymbol\varepsilon_t^\top)
=\boldsymbol\Phi^0$, and
$\max_{1\leq i\leq N}\mathrm E|\varepsilon_{it}|^{4+\delta}
\leq C_\varepsilon$, where
$\boldsymbol\tau^0=(\log\{(\sigma_1^0)^2\},\ldots,
\log\{(\sigma_N^0)^2\})^\top$ and
$\boldsymbol\Phi^0=\boldsymbol\Phi(\boldsymbol\tau^0)
=\operatorname{diag}\{(\sigma_1^0)^2,\ldots,(\sigma_N^0)^2\}$.
	\end{condition}
	
	\begin{condition}
		\label{ass:local_information}
		Let
		$\ddot Q_\eta(\boldsymbol\eta^0)
		=\left.\frac{\partial^2Q(\boldsymbol\eta)}
		{\partial\boldsymbol\eta_v\partial\boldsymbol\eta_v^\top}
		\right|_{\boldsymbol\eta=\boldsymbol\eta^0}$
		denote the population Hessian with respect to the vectorized coefficient
		and log-variance parameters. There exist constants
		$0<c_H<C_H<\infty$, independent of $N$ and $T$, such that,
		\begingroup\setlength{\abovedisplayskip}{4pt}\setlength{\belowdisplayskip}{4pt}\setlength{\abovedisplayshortskip}{2pt}\setlength{\belowdisplayshortskip}{2pt}
\begin{equation}
			c_H
			\leq
			\lambda_{\min}\!\left\{
			\mathbf D_{\eta,N}^{1/2}
			\ddot Q_\eta(\boldsymbol\eta^0)
			\mathbf D_{\eta,N}^{1/2}
			\right\}
			\leq
			\lambda_{\max}\!\left\{
			\mathbf D_{\eta,N}^{1/2}
			\ddot Q_\eta(\boldsymbol\eta^0)
			\mathbf D_{\eta,N}^{1/2}
			\right\}
			\leq C_H.
		\end{equation}
\endgroup
	\end{condition}

	Condition~\ref{ass:parameter_space_stability} ensures that the simultaneous
	system remains uniformly nonsingular and that the associated spatial
	multipliers are uniformly bounded. Such stability restrictions are
	standard in the asymptotic analysis of spatial QMLE
	\citep{lee2004asymptotic,lee2010estimation}.
	Condition~\ref{ass:true_parameter_interior}   ensures a regular interior
	parameter and stable  SVD normalization.
	Condition~\ref{ass:balanced_groups} prevents any local block from becoming
	asymptotically negligible relative to the full cross section.
Condition~\ref{ass:errors} assumes independence over time and across units for the errors. Its moment and independence assumptions control the linear and quadratic forms within each period and support a Lindeberg--Feller central limit theorem.
	Condition~\ref{ass:local_information} requires the scaled population
	Hessian to be uniformly positive definite and bounded. It therefore
	provides local identification and ensures the uniform invertibility needed
	for the quadratic expansion of the quasi-likelihood. Related curvature
	conditions are commonly imposed in the inferential theory for
	high-dimensional generalized factor models
	\citep{Wang2022GFM}.

	We now establish average consistency for the separately identified global
	and local coefficient components.
	
	\begin{theorem}
		\label{thm:separate_average_consistency}
		Suppose that Conditions~\ref{ass:feature_variation},
		\ref{ass:parameter_space_stability}, \ref{ass:balanced_groups},
		\ref{ass:errors}, and \ref{ass:local_information} hold, and let
		$N,T\to\infty$ with $N=o(T)$. Then, with probability approaching one,
		$Q_T$ admits a local minimizer
		$\widehat{\boldsymbol\eta}\in\mathcal H_N(r_g,r_l)$ satisfying
		\begingroup\setlength{\abovedisplayskip}{4pt}\setlength{\belowdisplayskip}{4pt}\setlength{\abovedisplayshortskip}{2pt}\setlength{\belowdisplayshortskip}{2pt}
\begin{equation}
			\|\widehat{\boldsymbol\eta}-\boldsymbol\eta^0\|_N^2
			=O_p(T^{-1}).
		\end{equation}
\endgroup
	\end{theorem}

	Theorem~\ref{thm:separate_average_consistency} implies
	$\|\widehat{\mathbf A}_g-\mathbf A_g^0\|_F^2=O_p(T^{-1})$ and
	$\|\widehat{\mathbf A}_{l,s}-\mathbf A_{l,s}^0\|_F^2=O_p(T^{-1})$ for every
	$s=1,\ldots,S$, while
	$N^{-1}\|\widehat{\boldsymbol\tau}-\boldsymbol\tau^0\|_2^2
	=O_p(T^{-1})$. Thus, the global and each local coefficient matrix converge
	at the root-$T$ rate in Frobenius norm, and the average squared error of the
	log-variance estimates is of order $T^{-1}$.
	
We next transfer the coefficient-space convergence rates to the normalized
factor representation. For this purpose, write
$
\boldsymbol\vartheta
=
(\boldsymbol\theta,\boldsymbol\tau)
=
\left(
\mathbf B,\mathbf C_g,
\{\mathbf U_s,\mathbf C_{l,s}\}_{s=1}^S,
\boldsymbol\tau
\right),
$
and let $\boldsymbol\vartheta_v$ denote the vector obtained by stacking these
blocks in the displayed order, with each matrix vectorized columnwise.
We use $\boldsymbol\vartheta^0$ and
$\widehat{\boldsymbol\vartheta}$ to denote the normalized true and estimated
factor representations, respectively, while a generic
$\boldsymbol\vartheta$ is not required to satisfy the normalization
conditions.

To account for the different dimensions of the loading and
characteristic-coefficient blocks, define the scaled factor norm by
	\begingroup\setlength{\abovedisplayskip}{4pt}\setlength{\belowdisplayskip}{4pt}\setlength{\abovedisplayshortskip}{2pt}\setlength{\belowdisplayshortskip}{2pt}
\begin{equation}
		\label{eq:factor_parameter_norm_main}
		\begin{aligned}
			\|\boldsymbol\vartheta-\boldsymbol\vartheta^0\|_{F,N}^2
			={}&
			N^{-1}\|\mathbf B-\mathbf B^0\|_F^2
			+\|\mathbf C_g-\mathbf C_g^0\|_F^2\\
			&+
			\sum_{s=1}^S
			\left\{
			N_s^{-1}\|\mathbf U_s-\mathbf U_s^0\|_F^2
			+\|\mathbf C_{l,s}-\mathbf C_{l,s}^0\|_F^2
			\right\}
			+
			N^{-1}\|\boldsymbol\tau-\boldsymbol\tau^0\|_2^2.
		\end{aligned}
	\end{equation}
\endgroup
	
	\begin{corollary}
		\label{cor:conditional_normalization_rate_transfer}
		Suppose that Conditions~\ref{ass:feature_variation}--
		\ref{ass:local_information} hold and $N=o(T)$. Let
		$\widehat{\boldsymbol\vartheta}$ be obtained from
		$\widehat{\boldsymbol\eta}$ by applying the SVD normalization in
		Proposition~\ref{prop:svd_internal_transformation} to the estimated
		coefficient matrices while leaving $\widehat{\boldsymbol\tau}$ unchanged.
		After paired sign alignment,
		\begingroup\setlength{\abovedisplayskip}{4pt}\setlength{\belowdisplayskip}{4pt}\setlength{\abovedisplayshortskip}{2pt}\setlength{\belowdisplayshortskip}{2pt}
\[
		\|\widehat{\boldsymbol\vartheta}
		-\boldsymbol\vartheta^0\|_{F,N}
		=O_p(T^{-1/2}).
		\]
\endgroup
	\end{corollary}
	
Corollary~\ref{cor:conditional_normalization_rate_transfer} shows that the coefficient-space convergence rate is preserved by the post-estimation SVD normalization. 
Specifically, the normalized global and local loading matrices, characteristic-coefficient matrices, and log-variance parameters
	are jointly consistent at the root-$T$ rate under the scaled factor norm.
	
	\subsection{Asymptotic Normality}
For individual asymptotic normality, consider a prespecified loading row or characteristic-coefficient column as $\{N,T\}\to\infty$.
Specifically, $1\leq i\leq N$
	for $\mathbf B$,
	$1\leq i\leq N_s$ for $\mathbf U_s$, $1\leq k\leq K$, and
	$1\leq s\leq S$. Define
	\begingroup\setlength{\abovedisplayskip}{4pt}\setlength{\belowdisplayskip}{4pt}\setlength{\abovedisplayshortskip}{2pt}\setlength{\belowdisplayshortskip}{2pt}
\begin{equation}
		\label{eq:main_factor_blocks}
		\boldsymbol\vartheta_m=
		\begin{cases}
			\mathbf B_{i\cdot}^{\top},&m=(B,i),\\
			\mathbf C_{g,\cdot k},&m=(C_g,k),\\
			\mathbf U_{s,i\cdot}^{\top},&m=(U,s,i),\\
			\mathbf C_{l,s,\cdot k},&m=(C_l,s,k).
		\end{cases}
	\end{equation}
\endgroup
Accordingly, $d_m=r_g$ for $m=(B,i)$ or $(C_g,k)$ and $d_m=r_l$ for $m=(U,s,i)$ or $(C_l,s,k)$. We use
$\widehat{\boldsymbol\vartheta}_m$ and
$\boldsymbol\vartheta_m^0$ for the corresponding blocks of the
sign-aligned local estimator in Corollary~\ref{cor:conditional_normalization_rate_transfer} and of the normalized truth.

For each fixed block $m$, let $\boldsymbol\Sigma_{m,N}$ denote the covariance matrix of the leading term in the first-order expansion of $\sqrt T(\widehat{\boldsymbol\vartheta}_m-\boldsymbol\vartheta_m^0)$.
Its explicit form is given  in
	Appendix~S3.
	
\begin{condition}
	\label{ass:block_covariance}
	For every fixed block $m$ in  \eqref{eq:main_factor_blocks}, there exists
	a positive-definite $d_m\times d_m$ matrix $\boldsymbol\Sigma_m$   with
	$0<c_\Sigma\le\lambda_{\min}(\boldsymbol\Sigma_m)
	\le\lambda_{\max}(\boldsymbol\Sigma_m)\le C_\Sigma<\infty$.
	As  $N,T\to\infty$, $\|\boldsymbol\Sigma_{m,N}-\boldsymbol\Sigma_m\|_F\to0$
	  for $m=(B,i)$ or $(U,s,i)$,
	$\|N\boldsymbol\Sigma_{m,N}-\boldsymbol\Sigma_m\|_F\to0$
	for $m=(C_g,k)$, and
	$\|N_s\boldsymbol\Sigma_{m,N}-\boldsymbol\Sigma_m\|_F\to0$
	for $m=(C_l,s,k)$.
\end{condition}

Following the fixed-index formulation of \citet{ouyang2026statistical},
Condition~\ref{ass:block_covariance} separately assumes that the
appropriately scaled covariance of each fixed block has a uniformly
positive definite limit. The scaling may differ between loading rows
and characteristic-coefficient columns.
	
\begin{theorem}
	\label{thm:individual_asymptotic_normality}
	Suppose that Conditions~\ref{ass:feature_variation}--
	\ref{ass:block_covariance} hold. If $N,T\to\infty$ with
	$N=o(T^{1/2})$, then
    for every fixed block $m$ in~\eqref{eq:main_factor_blocks},
 \begingroup\setlength{\abovedisplayskip}{4pt}\setlength{\belowdisplayskip}{4pt}\setlength{\abovedisplayshortskip}{2pt}\setlength{\belowdisplayshortskip}{2pt}
\[
	\boldsymbol\Sigma_{m,N}^{-1/2}\sqrt T
	\left(
	\widehat{\boldsymbol\vartheta}_m-\boldsymbol\vartheta_m^0
	\right)
	\xrightarrow{d}
	\mathcal N(\mathbf0,\mathbf I_{d_m}).
	\]
\endgroup
\end{theorem}

Theorem~\ref{thm:individual_asymptotic_normality}   gives the asymptotic normal distribution for each prespecified loading row or characteristic-coefficient column, standardized by its own covariance.

\begin{remark}
The average-consistency result only requires $N=o(T)$, whereas
Theorem~\ref{thm:individual_asymptotic_normality} is stated under the stronger condition $N=o(T^{1/2})$. The reason is that consistency only requires the estimation error to vanish under the scaled parameter norm, while fixed-block asymptotic normality further requires the remainder in the first-order linear representation to be negligible after covariance standardization. The covariance matrix $\boldsymbol\Sigma_{m,N}$ accounts for the different fluctuation scales across the loading and characteristic-coefficient blocks. Under this normalization, the Hessian-replacement remainder used in the proof is uniformly bounded by
$O_p(N/\sqrt T)$ over the four types of fixed blocks. Hence
$N=o(T^{1/2})$ is sufficient for this remainder to be $o_p(1)$ and yields a unified growth condition for Theorem~\ref{thm:individual_asymptotic_normality}.
\end{remark}

\subsection{Rank Selection}

The preceding results assume that the global and local ranks are known. We
now consider their joint selection using an information criterion. Let
$\boldsymbol r_0=(r_g,r_l)$ denote the true rank pair, and let
$\mathcal K=\{(k_g,k_l)\in\mathbb N^2:1\leq k_g,k_l\leq k_{\max}\}$ be the
candidate set, where $\max\{r_g,r_l\}\leq k_{\max}\leq K$ is fixed. For each
$\boldsymbol k=(k_g,k_l)\in\mathcal K$, let
$\mathcal V_N(\boldsymbol k)$ consist of
$(\boldsymbol\theta,\boldsymbol\tau)$, where
$\boldsymbol\theta=(\mathbf B,\mathbf C_g,
\{\mathbf U_s,\mathbf C_{l,s}\}_{s=1}^S)$,
$\boldsymbol\tau\in\mathbb R^N$,
$\mathbf B\in\mathbb R^{N\times k_g}$,
$\mathbf C_g\in\mathbb R^{k_g\times K}$,
$\mathbf U_s\in\mathbb R^{N_s\times k_l}$, and
$\mathbf C_{l,s}\in\mathbb R^{k_l\times K}$. Their coefficient products
and variance parameters are required to satisfy the compactness and
stability restrictions in
Condition~\ref{ass:parameter_space_stability}, with $(r_g,r_l)$ replaced by
$(k_g,k_l)$. Write
$\mathcal H_N(\boldsymbol k)=
\{\boldsymbol\eta(\boldsymbol\theta,\boldsymbol\tau):
(\boldsymbol\theta,\boldsymbol\tau)\in\mathcal V_N(\boldsymbol k)\}$.
		
For later comparison across candidate ranks, let
$\bar{\mathcal L}_T=2Q_T$ and $\bar{\mathcal L}=2Q$, and write
$\boldsymbol k\succeq\boldsymbol r_0$ if $k_g\geq r_g$ and $k_l\geq r_l$.
For an underfitted candidate
$\boldsymbol k\not\succeq\boldsymbol r_0$, let
$\boldsymbol\eta_{\boldsymbol k}^*$ denote a
 population minimizer of
$\bar{\mathcal L}$ over $\mathcal H_N(\boldsymbol k)$. For an exact-fit or
overfitted candidate $\boldsymbol k\succeq\boldsymbol r_0$, set
$\boldsymbol\eta_{\boldsymbol k}^*=\boldsymbol\eta^0$, which belongs to
$\mathcal H_N(\boldsymbol k)$ and minimizes the population criterion.

For each $\boldsymbol k\in\mathcal K$, let
$(\widehat{\boldsymbol\theta}_{\boldsymbol k},
\widehat{\boldsymbol\tau}_{\boldsymbol k})$ denote the rank-specific local
QMLE, and set
$\widehat{\boldsymbol\eta}_{\boldsymbol k}
=\boldsymbol\eta(\widehat{\boldsymbol\theta}_{\boldsymbol k},
\widehat{\boldsymbol\tau}_{\boldsymbol k})$.
We define the information criterion by
	\begingroup\setlength{\abovedisplayskip}{4pt}\setlength{\belowdisplayskip}{4pt}\setlength{\abovedisplayshortskip}{2pt}\setlength{\belowdisplayshortskip}{2pt}
\begin{equation}
		\label{eq:IC}
		\begin{aligned}
		\operatorname{IC}(\boldsymbol k)
			&=2Q_{T,F}(\widehat{\boldsymbol\theta}_{\boldsymbol k},
			\widehat{\boldsymbol\tau}_{\boldsymbol k})
		+(k_g+k_l)g(N,T),\\
		\widehat{\boldsymbol r}
		&\in\operatorname*{arg\,min}_{\boldsymbol k\in\mathcal K}
		\operatorname{IC}(\boldsymbol k).
		\end{aligned}
	\end{equation}
\endgroup
Algorithm~\ref{al2} implements rank selection using the numerical estimates from Algorithm~\ref{al1}.
{For exact-fit and overfitted candidates, use the local
minimizer near $\boldsymbol\eta^0$ constructed in
Appendix~S4.}
	
{
\begin{condition}
\label{ass:rank_underfit}
There is a constant $c_{\min}>0$, independent of
$\boldsymbol k,N,T$, such that every underfitted candidate
$\boldsymbol k\not\succeq\boldsymbol r_0$ satisfies
\begingroup\setlength{\abovedisplayskip}{4pt}\setlength{\belowdisplayskip}{4pt}\setlength{\abovedisplayshortskip}{2pt}\setlength{\belowdisplayshortskip}{2pt}
\[
\bar{\mathcal L}(\boldsymbol\eta_{\boldsymbol k}^*)
-\bar{\mathcal L}(\boldsymbol\eta^0)
\ge\frac{c_{\min}}N.
\]
\endgroup
\end{condition}

Condition~\ref{ass:rank_underfit} requires the minimum population
criterion under every underfitted rank specification to exceed
$\bar{\mathcal L}(\boldsymbol\eta^0)$ by at least $c_{\min}/N$.
The minimization is joint over the coefficient matrices and the
log-variance vector in $\mathcal H_N(\boldsymbol k)$, with
$\boldsymbol\eta_{\boldsymbol k}^*$ denoting a population minimizer.
For exact-fit and overfitted candidates, local quadratic growth
around $\boldsymbol\eta^0$ follows from
Condition~\ref{ass:local_information}.

\begin{remark}[Verification of Condition~\ref{ass:rank_underfit}]
Proposition~5 in Section~S4.2 of the supplementary material gives a sufficient condition for Condition~\ref{ass:rank_underfit}. It requires the covariance matrices at each underfitted population minimizer to remain separated from the true covariance matrices, as measured by their time-averaged squared Frobenius distance. Under the stability bounds, this gives the required population loss of order $N^{-1}$. Section~S5.6 provides numerical evidence for this loss in the designs considered.
\end{remark}
}

	\begin{theorem}
		\label{th3}
		Let $\widehat{\boldsymbol r}$ be defined by~\eqref{eq:IC}. Suppose that
		Condition~\ref{ass:feature_variation},
		Conditions~\ref{ass:parameter_space_stability}--\ref{ass:local_information},
		and Condition~\ref{ass:rank_underfit}
		hold. If $(N,T)\to\infty$, $N=o(T)$, $Tg(N,T)\to\infty$, and
		$Ng(N,T)\to0$, then
		$\Pr(\widehat{\boldsymbol r}=\boldsymbol r_0)\to1$.
	\end{theorem}
	
The restrictions $Tg(N,T)\to\infty$ and $Ng(N,T)\to0$ balance the effects
of overfitting and underfitting. Specifically, $Tg(N,T)\to\infty$ ensures
that the rank penalty dominates the $O_p(T^{-1})$ reduction in the sample
criterion due to overfitting, whereas $Ng(N,T)\to0$ makes the penalty
asymptotically negligible relative to the order-$N^{-1}$ population loss
from underfitting. We set
$g(N,T)=c_g\log(NT)/\sqrt{NT}$, which satisfies these restrictions when
$N\log^2(NT)/T\to0$.
In the simulation study, we use $c_g=0.05$ as the baseline constant, which gives reasonable rank-selection accuracy in the designs considered. To assess robustness to this choice, we also conduct a sensitivity analysis over a range of constants in Appendix~S5.5.

\IncMargin{1em}
\begin{algorithm}[h!]
	\linespread{0.8}\selectfont
	\caption{Rank selection: minimization of $\operatorname{IC}(k_g,k_l)$ \label{al2}}
	\SetAlgoNoLine
	\SetKwInOut{Input}{\textbf{Input}}
	\SetKwInOut{Output}{\textbf{Output}}
	
	\Input{Maximum rank: $k_{\max}$.\\}
	\Output{$\widehat r_g,\widehat r_l$.\\}
	\BlankLine
	
	\For{$k_g=1,\ldots,k_{\max}$ and $k_l=1,\ldots,k_{\max}$}{
		Run Algorithm~\ref{al1} with ranks $(k_g,k_l)$ and calculate
		$\operatorname{IC}(k_g,k_l)$\;
	}
	Select $\widehat r_g$ and $\widehat r_l$ that minimize
	$\operatorname{IC}(k_g,k_l)$ in \eqref{eq:IC}\;
	{\textbf{return}} $\widehat r_g,\widehat r_l$\;
\end{algorithm}
\DecMargin{1em}

\section{Simulation Studies} \label{sec3}

{We use Monte Carlo simulations to evaluate the finite-sample performance of QMLE and the accuracy of global and local rank selection.}
Two distinct simulation environments are considered.
The first model adopts independent, idiosyncratic errors with various tail behaviors, whereas the second incorporates cross-sectional dependence and heteroskedasticity commonly observed in economic and financial applications.

\subsection{Two Simulation Settings}

We construct the data-generating processes based on the model specified in~\eqref{eq:global_2_compact}.
{To obtain coefficient matrices of the specified ranks and a stable simultaneous system, we generate the loading matrices and errors as follows.}

\noindent \textbf{Case 1.}
We first generate matrices $\B_0$ and $\mathbf C_{g0}$.
Let $\mathbf Q_B\in\mathbb R^{N\times r_g}$ and $\mathbf Q_{C,g}\in\mathbb R^{K\times r_g}$ be orthonormal bases obtained from independent standard Gaussian matrices.
{We consider $r_g,r_l\in\{1,2\}$. For $r_g=1$, let $\mathbf D_g$ be the $1\times1$ matrix with entry 1; for $r_g=2$, let $\mathbf D_g=\operatorname{diag}(1,0.5)$. We construct} $\B_0=\mathbf Q_B\mathbf D_g^{1/2}$ and $\mathbf C_{g0}=\mathbf D_g^{1/2}\mathbf Q_{C,g}^{\top}$.
Thus, the global characteristic coefficient matrix $\mathbf A_{g0}=\B_0\mathbf C_{g0}$ has rank $r_g$.
Similarly, for group $s$, we independently generate orthonormal bases $\mathbf Q_{U,s}\in\mathbb R^{N_s\times r_l}$ and $\mathbf Q_{C,l,s}\in\mathbb R^{K\times r_l}$.
{Define $\mathbf D_l$ in the same way using $r_l$. We construct} $\U_{s0}=\mathbf Q_{U,s}\mathbf D_l^{1/2}$ and $\mathbf C_{l,s0}=\mathbf D_l^{1/2}\mathbf Q_{C,l,s}^{\top}$, so the local characteristic coefficient matrix $\mathbf A_{l,s0}=\U_{s0}\mathbf C_{l,s0}$ has rank $r_l$.
At each $t$, we generate a characteristic matrix
$\mathbf Z_{t-1}\in\mathbb R^{K\times N}$ whose entries are independent
draws from $U[-1,1]$, where $K$ denotes the number of characteristics, with $K=3$ in the simulations. The global component uses $\mathbf Z_{t-1}$,
and the local component for group $s$ uses the corresponding columns,
denoted by $\mathbf Z_{s,t-1}\in\mathbb R^{K\times N_s}$.
The resulting time-varying interaction matrix is
$
\mathbf M_{0t}=\mathbf A_{g0}\mathbf Z_{t-1}+\operatorname{bdiag}\{\mathbf A_{l,10}\mathbf Z_{1,t-1},\ldots,\mathbf A_{l,S0}\mathbf Z_{S,t-1}\}.
$
{To set the scale of the coefficients, we generate a separate characteristic path of length $T$ and multiply $\B_0$, $\mathbf C_{g0}$, $\U_0$, and $\mathbf C_{l0}$ by a common constant so that $\max_t\|\mathbf M_{0t}\|_2=0.8$ along this path,}
and generate $\Y_t=(\mathbf I_N-\mathbf M_{0t})^{-1}
\boldsymbol\varepsilon_t$, $t=1,\ldots,T$.

For the noise term $\boldsymbol\varepsilon_t$, we first draw
$s_i\stackrel{\mathrm{iid}}{\sim}U[0.5,1]$ and set
$\boldsymbol\Phi_0=\operatorname{diag}(s_1,\ldots,s_N)$.
To examine sensitivity to the error distribution, we generate
$\boldsymbol\varepsilon_t=\boldsymbol\Phi_0^{1/2}\mathbf z_t$
under three specifications:
(1) the elements of $\mathbf z_t$ are independent standard normal variables;
(2) the elements of $\mathbf z_t$ are independent $t(5)$ variables
divided by $\sqrt{5/3}$; and
(3) the elements of $\mathbf z_t$ are independent draws from
$\{0.3N(0,1)+0.7N(0,4)\}/\sqrt{3.1}$.
In all cases, given $\boldsymbol\Phi_0$, the errors have mean zero
and covariance matrix $\boldsymbol\Phi_0$.

\noindent \textbf{Case 2.}
The second experiment examines estimation under blockwise cross-sectional dependence and heteroskedasticity.
The coefficient and characteristic processes are generated as in Case 1, but the covariance matrix $\boldsymbol\Phi_0$ is different.
Following the design in \citet{fan2015power}, we initialize $\mathbf R=\mathbf I_N$ and partition the first $4\lfloor N/4\rfloor$ units into $\lfloor N/4\rfloor$ blocks of size four.
We randomly select $\lfloor N^{0.3}\rfloor$ of these blocks and replace each selected block by the correlation matrix whose $(l,k)$ entry is $0.2^{|l-k|}$ for $1\leq l,k\leq4$.
Any remaining units are uncorrelated with one another and with all other units.
To incorporate cross-sectional heteroskedasticity, we draw $s_i\stackrel{\mathrm{iid}}{\sim}U[1,1.5]$, define $\mathbf D=\operatorname{diag}(s_1,\ldots,s_N)$, and set $\boldsymbol\Phi_0=\mathbf D^{1/2}\mathbf R\mathbf D^{1/2}$.
This design assesses robustness
when the errors exhibit cross-sectional correlation and heteroskedasticity.

{
We set the number of groups to $S=3$ with equal group sizes $N_1=N_2=N_3=N/3$.
The true ranks for global and local factors, $(r_g, r_l)$, vary across the set $\{1, 2\}$, yielding four distinct combinations. We consider a range of sample sizes with $N \in \{60, 90, 120\}$ and $T \in \{1000, 1500, 2000\}$, resulting in nine $(N, T)$ pairs. These settings are similar to the standard data sets used in economics and include combinations similar to the cross-sectional and time-series dimensions of our industry portfolio application. For each setting, we replicate the experiment 200 times, keeping the data-generating coefficient matrices and error variances fixed across replications.

For the QMLE procedure in Algorithm~\ref{al1}, we use the normalized objective $\mathcal L_{T,c}/T$ and set $M=3$, $\chi=10^{-2}$, $K_{\mathrm{in}}=100$, $\varpi=0.5$, and $\xi=10^{-6}$. 
For each random start, we generate initial values by drawing auxiliary loading and characteristic-coefficient entries independently from $N(0,1)$ and $U[-1,1]$, respectively. An SVD of their products, followed by a common rescaling, yields $\mathbf B^{(0)}$, $\mathbf C_g^{(0)}$ and, for each group, $\mathbf U_s^{(0)}$, $\mathbf C_{l,s}^{(0)}$. Among the converged estimates, we retain the one with the highest concentrated quasi-log-likelihood.
\par}

\subsection{Simulation Results for QMLE Estimation}

{To quantify the estimation accuracy, we report the means (Mean) and standard deviations (SD) of the estimation errors across 200 replications.} We use the Frobenius norm for matrices and the Euclidean norm for vectors, scaling each by its respective dimensions to facilitate comparison. 
Specifically, the estimation errors for the model parameters are computed as follows: 
$\|\widehat{\mathbf A}_g-\mathbf A_{g0}\|_F/\sqrt{NK}$,
$\|\widehat{\B}-\B_0\|_F/\sqrt{Nr_g}$,
$\|\widehat{\mathbf C}_g-\mathbf C_{g0}\|_F/\sqrt{r_gK}$,
$\|\widehat{\mathbf A}_l-\mathbf A_{l0}\|_F/\sqrt{NK}$,
$\|\widehat{\U}-\U_0\|_F/\sqrt{Nr_l}$,
$\|\widehat{\mathbf C}_l-\mathbf C_{l0}\|_F/\sqrt{Sr_lK}$,
and
$\|\widehat{\boldsymbol\sigma}^{2}-\boldsymbol\sigma_0^{2}\|_2/\sqrt N$.
Here, $\mathbf B_0$, $\mathbf C_{g0}$, $\mathbf U_0$, and $\mathbf C_{l0}$ denote the corresponding normalized true parameters.
We also report the response-fit error $\{\frac{1}{NT}\sum_{t=1}^T\|(\widehat{\mathbf M}_t-\mathbf M_{0t})\Y_t\|_2^2\}^{1/2}$, {reported in the columns labeled Fit error.}
In these metrics, lower values indicate more accurate estimation.

\begin{table}[h!]
\caption{
Means and standard deviations (in parentheses) for Case 1 under normal errors.}
\label{case1_norm}

\vspace{-0.1cm}
{\centering
\footnotesize{
\setlength{\tabcolsep}{0.25em}
\renewcommand{\arraystretch}{1}
\resizebox{\textwidth}{!}{
\begin{tabular}{cccccccccc}
\toprule
$N$ & $T$ & $\mathbf A_g$ & $\B$ & $\mathbf C_g$ & $\mathbf A_l$ & $\U$ & $\mathbf C_l$ & $\pmb\sigma^2$ & {Fit error} \\
\hline
& & \multicolumn{8}{c}{Panel A: $r_g=r_l=1$} \\
\hline
60 & 1000 & 0.0046(0.0004) & 0.0438(0.0040) & 0.0099(0.0035) & 0.0081(0.0008) & 0.0560(0.0058) & 0.0201(0.0049) & 0.0350(0.0032) & 0.0404(0.0027) \\
 & 1500 & 0.0038(0.0004) & 0.0364(0.0034) & 0.0078(0.0029) & 0.0066(0.0006) & 0.0463(0.0045) & 0.0161(0.0040) & 0.0286(0.0027) & 0.0331(0.0020) \\
 & 2000 & 0.0033(0.0003) & 0.0316(0.0029) & 0.0063(0.0026) & 0.0057(0.0005) & 0.0404(0.0037) & 0.0135(0.0035) & 0.0247(0.0024) & 0.0285(0.0017) \\
90 & 1000 & 0.0038(0.0003) & 0.0442(0.0036) & 0.0095(0.0028) & 0.0067(0.0005) & 0.0573(0.0042) & 0.0191(0.0046) & 0.0341(0.0028) & 0.0399(0.0022) \\
 & 1500 & 0.0031(0.0002) & 0.0364(0.0030) & 0.0071(0.0021) & 0.0055(0.0004) & 0.0475(0.0034) & 0.0143(0.0033) & 0.0278(0.0022) & 0.0323(0.0017) \\
 & 2000 & 0.0027(0.0002) & 0.0317(0.0025) & 0.0058(0.0023) & 0.0047(0.0004) & 0.0412(0.0033) & 0.0117(0.0027) & 0.0242(0.0019) & 0.0278(0.0015) \\
120 & 1000 & 0.0032(0.0002) & 0.0395(0.0028) & 0.0082(0.0022) & 0.0056(0.0004) & 0.0516(0.0032) & 0.0160(0.0036) & 0.0345(0.0022) & 0.0395(0.0019) \\
 & 1500 & 0.0026(0.0002) & 0.0330(0.0022) & 0.0059(0.0019) & 0.0045(0.0003) & 0.0427(0.0028) & 0.0122(0.0030) & 0.0281(0.0019) & 0.0321(0.0013) \\
 & 2000 & 0.0022(0.0001) & 0.0290(0.0019) & 0.0047(0.0016) & 0.0039(0.0003) & 0.0375(0.0025) & 0.0103(0.0026) & 0.0242(0.0018) & 0.0278(0.0013) \\
\hline
& & \multicolumn{8}{c}{Panel B: $r_g=1,\ r_l=2$} \\
\hline
60 & 1000 & 0.0047(0.0005) & 0.0447(0.0041) & 0.0106(0.0043) & 0.0110(0.0008) & 0.0652(0.0051) & 0.0323(0.0088) & 0.0343(0.0033) & 0.0490(0.0027) \\
 & 1500 & 0.0039(0.0004) & 0.0368(0.0035) & 0.0082(0.0035) & 0.0089(0.0006) & 0.0543(0.0039) & 0.0241(0.0068) & 0.0279(0.0027) & 0.0399(0.0020) \\
 & 2000 & 0.0033(0.0003) & 0.0320(0.0032) & 0.0069(0.0030) & 0.0077(0.0006) & 0.0475(0.0037) & 0.0203(0.0054) & 0.0241(0.0023) & 0.0345(0.0019) \\
90 & 1000 & 0.0037(0.0003) & 0.0417(0.0036) & 0.0091(0.0030) & 0.0088(0.0005) & 0.0607(0.0039) & 0.0284(0.0067) & 0.0344(0.0027) & 0.0488(0.0022) \\
 & 1500 & 0.0030(0.0003) & 0.0342(0.0031) & 0.0069(0.0023) & 0.0071(0.0004) & 0.0506(0.0036) & 0.0209(0.0057) & 0.0279(0.0021) & 0.0395(0.0017) \\
 & 2000 & 0.0026(0.0002) & 0.0298(0.0025) & 0.0057(0.0022) & 0.0061(0.0003) & 0.0440(0.0024) & 0.0168(0.0039) & 0.0244(0.0018) & 0.0340(0.0014) \\
120 & 1000 & 0.0034(0.0002) & 0.0431(0.0030) & 0.0101(0.0028) & 0.0078(0.0004) & 0.0623(0.0037) & 0.0298(0.0079) & 0.0345(0.0020) & 0.0488(0.0018) \\
 & 1500 & 0.0027(0.0002) & 0.0361(0.0024) & 0.0071(0.0023) & 0.0063(0.0003) & 0.0520(0.0029) & 0.0210(0.0051) & 0.0280(0.0019) & 0.0395(0.0014) \\
 & 2000 & 0.0024(0.0001) & 0.0316(0.0020) & 0.0056(0.0019) & 0.0055(0.0003) & 0.0461(0.0026) & 0.0172(0.0049) & 0.0241(0.0018) & 0.0342(0.0012) \\
\hline
& & \multicolumn{8}{c}{Panel C: $r_g=2,\ r_l=1$} \\
\hline
60 & 1000 & 0.0063(0.0004) & 0.0515(0.0034) & 0.0162(0.0043) & 0.0085(0.0007) & 0.0612(0.0058) & 0.0255(0.0064) & 0.0350(0.0031) & 0.0495(0.0024) \\
 & 1500 & 0.0052(0.0004) & 0.0432(0.0032) & 0.0123(0.0037) & 0.0069(0.0006) & 0.0505(0.0043) & 0.0204(0.0053) & 0.0285(0.0028) & 0.0402(0.0020) \\
 & 2000 & 0.0044(0.0003) & 0.0378(0.0028) & 0.0100(0.0032) & 0.0059(0.0005) & 0.0436(0.0039) & 0.0169(0.0043) & 0.0246(0.0023) & 0.0347(0.0019) \\
90 & 1000 & 0.0051(0.0003) & 0.0456(0.0027) & 0.0131(0.0039) & 0.0067(0.0005) & 0.0537(0.0038) & 0.0185(0.0046) & 0.0339(0.0027) & 0.0483(0.0020) \\
 & 1500 & 0.0041(0.0002) & 0.0382(0.0021) & 0.0096(0.0029) & 0.0054(0.0004) & 0.0441(0.0034) & 0.0146(0.0036) & 0.0275(0.0021) & 0.0392(0.0017) \\
 & 2000 & 0.0035(0.0002) & 0.0334(0.0018) & 0.0076(0.0024) & 0.0047(0.0003) & 0.0380(0.0028) & 0.0123(0.0032) & 0.0240(0.0019) & 0.0337(0.0015) \\
120 & 1000 & 0.0043(0.0002) & 0.0435(0.0021) & 0.0124(0.0029) & 0.0057(0.0004) & 0.0521(0.0035) & 0.0173(0.0038) & 0.0341(0.0021) & 0.0479(0.0017) \\
 & 1500 & 0.0035(0.0002) & 0.0367(0.0017) & 0.0091(0.0020) & 0.0046(0.0003) & 0.0428(0.0030) & 0.0129(0.0029) & 0.0277(0.0018) & 0.0389(0.0014) \\
 & 2000 & 0.0030(0.0001) & 0.0324(0.0016) & 0.0071(0.0016) & 0.0039(0.0003) & 0.0374(0.0025) & 0.0105(0.0026) & 0.0239(0.0017) & 0.0336(0.0012) \\
\hline
& & \multicolumn{8}{c}{Panel D: $r_g=r_l=2$} \\
\hline
60 & 1000 & 0.0068(0.0005) & 0.0545(0.0037) & 0.0182(0.0054) & 0.0120(0.0008) & 0.0720(0.0058) & 0.0391(0.0100) & 0.0332(0.0033) & 0.0557(0.0026) \\
 & 1500 & 0.0055(0.0004) & 0.0455(0.0032) & 0.0134(0.0048) & 0.0096(0.0006) & 0.0593(0.0047) & 0.0292(0.0077) & 0.0271(0.0028) & 0.0451(0.0021) \\
 & 2000 & 0.0047(0.0003) & 0.0398(0.0029) & 0.0105(0.0038) & 0.0083(0.0006) & 0.0519(0.0041) & 0.0239(0.0063) & 0.0233(0.0023) & 0.0388(0.0018) \\
90 & 1000 & 0.0055(0.0003) & 0.0499(0.0029) & 0.0160(0.0044) & 0.0096(0.0006) & 0.0664(0.0045) & 0.0327(0.0080) & 0.0349(0.0027) & 0.0568(0.0022) \\
 & 1500 & 0.0044(0.0002) & 0.0419(0.0024) & 0.0119(0.0033) & 0.0077(0.0004) & 0.0552(0.0036) & 0.0236(0.0061) & 0.0284(0.0022) & 0.0461(0.0017) \\
 & 2000 & 0.0038(0.0002) & 0.0365(0.0021) & 0.0091(0.0026) & 0.0066(0.0004) & 0.0479(0.0028) & 0.0193(0.0043) & 0.0247(0.0020) & 0.0395(0.0014) \\
120 & 1000 & 0.0047(0.0002) & 0.0482(0.0023) & 0.0158(0.0036) & 0.0084(0.0004) & 0.0652(0.0039) & 0.0323(0.0084) & 0.0334(0.0020) & 0.0554(0.0018) \\
 & 1500 & 0.0038(0.0002) & 0.0406(0.0021) & 0.0116(0.0028) & 0.0068(0.0004) & 0.0544(0.0032) & 0.0231(0.0062) & 0.0272(0.0018) & 0.0449(0.0014) \\
 & 2000 & 0.0033(0.0002) & 0.0360(0.0018) & 0.0091(0.0022) & 0.0058(0.0003) & 0.0476(0.0027) & 0.0184(0.0045) & 0.0234(0.0017) & 0.0386(0.0013) \\
\bottomrule
\end{tabular}}}
\par}
\vspace{0.12cm}
\footnotesize{Notes: 
This table reports mean estimation errors and standard deviations (in parentheses) for Case 1 under normal errors. Columns report errors for the global parameters $\mathbf A_g$, $\mathbf B$, $\mathbf C_g$, local parameters $\mathbf A_l$, $\mathbf U$, $\mathbf C_l$, error variances $\boldsymbol\sigma^2$, and {response-fit error (Fit error)}. 
The number of global ($r_g$) and local ($r_l$) factors was chosen from $\{1, 2\}$, yielding four combinations detailed in Panels A to D. 
}
\end{table}

\begin{table}[h!]
\caption{
Means and standard deviations (in parentheses) for Case 2 under normal errors.}
\label{case2_norm}
\vspace{-0.1cm}
{\centering
\footnotesize{
\setlength{\tabcolsep}{0.25em}
\renewcommand{\arraystretch}{1}
\resizebox{\textwidth}{!}{
\begin{tabular}{cccccccccc}
\toprule
$N$ & $T$ & $\mathbf A_g$ & $\B$ & $\mathbf C_g$ & $\mathbf A_l$ & $\U$ & $\mathbf C_l$ & $\pmb\sigma^2$ & {Fit error} \\
\hline
& & \multicolumn{8}{c}{Panel A: $r_g=r_l=1$} \\
\hline
60 & 1000 & 0.0047(0.0004) & 0.0442(0.0043) & 0.0100(0.0034) & 0.0081(0.0008) & 0.0562(0.0058) & 0.0202(0.0053) & 0.0571(0.0052) & 0.0521(0.0035) \\
 & 1500 & 0.0038(0.0003) & 0.0363(0.0035) & 0.0073(0.0032) & 0.0066(0.0006) & 0.0465(0.0046) & 0.0160(0.0037) & 0.0467(0.0043) & 0.0424(0.0027) \\
 & 2000 & 0.0032(0.0003) & 0.0314(0.0029) & 0.0060(0.0025) & 0.0057(0.0006) & 0.0405(0.0042) & 0.0133(0.0033) & 0.0406(0.0036) & 0.0365(0.0025) \\
90 & 1000 & 0.0039(0.0003) & 0.0444(0.0036) & 0.0097(0.0029) & 0.0068(0.0006) & 0.0579(0.0051) & 0.0193(0.0045) & 0.0560(0.0043) & 0.0521(0.0030) \\
 & 1500 & 0.0031(0.0003) & 0.0369(0.0029) & 0.0072(0.0024) & 0.0055(0.0004) & 0.0477(0.0038) & 0.0147(0.0036) & 0.0455(0.0036) & 0.0422(0.0022) \\
 & 2000 & 0.0027(0.0002) & 0.0320(0.0025) & 0.0058(0.0019) & 0.0048(0.0004) & 0.0418(0.0034) & 0.0122(0.0031) & 0.0394(0.0031) & 0.0364(0.0019) \\
120 & 1000 & 0.0032(0.0002) & 0.0402(0.0026) & 0.0081(0.0024) & 0.0057(0.0004) & 0.0524(0.0034) & 0.0164(0.0036) & 0.0565(0.0033) & 0.0514(0.0022) \\
 & 1500 & 0.0026(0.0002) & 0.0332(0.0021) & 0.0058(0.0020) & 0.0046(0.0003) & 0.0433(0.0030) & 0.0121(0.0030) & 0.0461(0.0028) & 0.0416(0.0019) \\
 & 2000 & 0.0022(0.0001) & 0.0290(0.0018) & 0.0046(0.0016) & 0.0039(0.0003) & 0.0377(0.0025) & 0.0100(0.0024) & 0.0399(0.0028) & 0.0359(0.0016) \\
\hline
& & \multicolumn{8}{c}{Panel B: $r_g=1,\ r_l=2$} \\
\hline
60 & 1000 & 0.0047(0.0005) & 0.0448(0.0045) & 0.0105(0.0036) & 0.0110(0.0009) & 0.0657(0.0058) & 0.0329(0.0092) & 0.0564(0.0053) & 0.0633(0.0038) \\
 & 1500 & 0.0038(0.0004) & 0.0369(0.0036) & 0.0080(0.0032) & 0.0089(0.0006) & 0.0542(0.0040) & 0.0247(0.0059) & 0.0461(0.0044) & 0.0514(0.0029) \\
 & 2000 & 0.0033(0.0003) & 0.0319(0.0030) & 0.0067(0.0027) & 0.0077(0.0005) & 0.0471(0.0034) & 0.0203(0.0047) & 0.0400(0.0037) & 0.0443(0.0025) \\
90 & 1000 & 0.0038(0.0003) & 0.0420(0.0036) & 0.0092(0.0030) & 0.0089(0.0005) & 0.0610(0.0039) & 0.0286(0.0077) & 0.0565(0.0044) & 0.0634(0.0032) \\
 & 1500 & 0.0031(0.0003) & 0.0346(0.0028) & 0.0070(0.0025) & 0.0072(0.0004) & 0.0503(0.0031) & 0.0203(0.0046) & 0.0458(0.0036) & 0.0513(0.0024) \\
 & 2000 & 0.0026(0.0002) & 0.0299(0.0023) & 0.0056(0.0020) & 0.0062(0.0003) & 0.0443(0.0026) & 0.0170(0.0040) & 0.0396(0.0030) & 0.0443(0.0021) \\
120 & 1000 & 0.0034(0.0002) & 0.0436(0.0028) & 0.0099(0.0027) & 0.0079(0.0004) & 0.0626(0.0034) & 0.0301(0.0071) & 0.0566(0.0033) & 0.0633(0.0023) \\
 & 1500 & 0.0027(0.0002) & 0.0362(0.0024) & 0.0071(0.0023) & 0.0064(0.0003) & 0.0526(0.0030) & 0.0222(0.0054) & 0.0462(0.0029) & 0.0514(0.0020) \\
 & 2000 & 0.0024(0.0002) & 0.0317(0.0020) & 0.0058(0.0020) & 0.0055(0.0003) & 0.0460(0.0023) & 0.0176(0.0040) & 0.0400(0.0029) & 0.0442(0.0016) \\
\hline
& & \multicolumn{8}{c}{Panel C: $r_g=2,\ r_l=1$} \\
\hline
60 & 1000 & 0.0064(0.0004) & 0.0517(0.0033) & 0.0168(0.0056) & 0.0086(0.0008) & 0.0614(0.0058) & 0.0262(0.0075) & 0.0572(0.0053) & 0.0639(0.0035) \\
 & 1500 & 0.0051(0.0003) & 0.0432(0.0029) & 0.0123(0.0038) & 0.0069(0.0007) & 0.0504(0.0051) & 0.0207(0.0058) & 0.0466(0.0043) & 0.0518(0.0028) \\
 & 2000 & 0.0044(0.0003) & 0.0378(0.0023) & 0.0100(0.0031) & 0.0059(0.0006) & 0.0437(0.0044) & 0.0174(0.0046) & 0.0406(0.0037) & 0.0447(0.0023) \\
90 & 1000 & 0.0051(0.0003) & 0.0458(0.0027) & 0.0130(0.0043) & 0.0068(0.0005) & 0.0537(0.0043) & 0.0184(0.0043) & 0.0557(0.0043) & 0.0629(0.0030) \\
 & 1500 & 0.0041(0.0002) & 0.0382(0.0022) & 0.0092(0.0027) & 0.0055(0.0004) & 0.0443(0.0032) & 0.0143(0.0034) & 0.0453(0.0036) & 0.0509(0.0023) \\
 & 2000 & 0.0035(0.0002) & 0.0335(0.0019) & 0.0074(0.0020) & 0.0047(0.0004) & 0.0384(0.0029) & 0.0121(0.0029) & 0.0392(0.0031) & 0.0439(0.0019) \\
120 & 1000 & 0.0044(0.0002) & 0.0441(0.0021) & 0.0127(0.0033) & 0.0057(0.0004) & 0.0522(0.0034) & 0.0173(0.0041) & 0.0562(0.0032) & 0.0624(0.0024) \\
 & 1500 & 0.0035(0.0002) & 0.0369(0.0017) & 0.0091(0.0021) & 0.0046(0.0003) & 0.0433(0.0028) & 0.0133(0.0030) & 0.0459(0.0029) & 0.0507(0.0020) \\
 & 2000 & 0.0030(0.0001) & 0.0326(0.0014) & 0.0074(0.0021) & 0.0040(0.0003) & 0.0379(0.0024) & 0.0109(0.0026) & 0.0397(0.0028) & 0.0438(0.0017) \\
\hline
& & \multicolumn{8}{c}{Panel D: $r_g=r_l=2$} \\
\hline
60 & 1000 & 0.0068(0.0005) & 0.0546(0.0037) & 0.0183(0.0058) & 0.0121(0.0009) & 0.0728(0.0065) & 0.0395(0.0106) & 0.0554(0.0050) & 0.0727(0.0035) \\
 & 1500 & 0.0055(0.0004) & 0.0457(0.0033) & 0.0135(0.0043) & 0.0097(0.0007) & 0.0600(0.0050) & 0.0300(0.0081) & 0.0452(0.0042) & 0.0589(0.0029) \\
 & 2000 & 0.0048(0.0003) & 0.0399(0.0028) & 0.0107(0.0035) & 0.0084(0.0007) & 0.0524(0.0043) & 0.0247(0.0069) & 0.0393(0.0038) & 0.0508(0.0026) \\
90 & 1000 & 0.0055(0.0003) & 0.0503(0.0029) & 0.0162(0.0047) & 0.0097(0.0006) & 0.0672(0.0047) & 0.0347(0.0090) & 0.0568(0.0044) & 0.0737(0.0030) \\
 & 1500 & 0.0044(0.0002) & 0.0421(0.0022) & 0.0116(0.0032) & 0.0078(0.0004) & 0.0555(0.0035) & 0.0248(0.0066) & 0.0461(0.0036) & 0.0595(0.0022) \\
 & 2000 & 0.0038(0.0002) & 0.0367(0.0020) & 0.0090(0.0025) & 0.0067(0.0004) & 0.0484(0.0028) & 0.0199(0.0050) & 0.0399(0.0030) & 0.0512(0.0020) \\
120 & 1000 & 0.0048(0.0002) & 0.0487(0.0022) & 0.0157(0.0036) & 0.0085(0.0004) & 0.0655(0.0036) & 0.0325(0.0081) & 0.0556(0.0031) & 0.0725(0.0021) \\
 & 1500 & 0.0039(0.0002) & 0.0410(0.0018) & 0.0112(0.0025) & 0.0069(0.0003) & 0.0551(0.0030) & 0.0242(0.0062) & 0.0454(0.0028) & 0.0587(0.0020) \\
 & 2000 & 0.0033(0.0002) & 0.0363(0.0017) & 0.0091(0.0023) & 0.0059(0.0003) & 0.0482(0.0025) & 0.0192(0.0048) & 0.0393(0.0027) & 0.0506(0.0017) \\
\bottomrule
\end{tabular}}
}
\par}
\vspace{0.12cm}
\footnotesize{Notes: 
This table reports mean estimation errors and standard deviations (in parentheses) for Case 2 under normal errors.  The table format follows Table \ref{case1_norm}. 
 }
\end{table}

Tables~\ref{case1_norm} and~\ref{case2_norm} summarize the simulation results for Cases 1 and 2 under normal errors, respectively. {Three patterns emerge.} {First, for fixed $N$ and ranks, the mean estimation errors and response-fit errors decrease as $T$ increases, while their standard deviations generally decline. }
Second, comparing the two cases, coefficient matrix errors are similar, whereas variance estimation and response-fit errors are larger under the dependent errors in Case 2. Nevertheless, both errors decrease with $T$. 
Third, for fixed $T$ and ranks, increasing $N$ generally reduces coefficient matrix errors, although its effects on individual factor parameter errors are less uniform.

The supplementary material examines robustness to non-Gaussian errors (Section~S5.1), unequal local ranks (Section~S5.2), sparse portfolio weights (Section~S5.3), and larger cross-sectional dimensions (Section~S5.4). Overall, the estimation errors exhibit broadly similar patterns across these settings, supporting the robustness of the estimator's finite-sample performance.

\subsection{Simulation Results for Rank Selection}

We then evaluate the performance of the IC proposed in~\eqref{eq:IC} for determining the numbers of global and local factors. 
{Table~\ref{rank_norm} reports the probabilities of correctly selecting each rank and both ranks jointly using the IC in Eq.~(\ref{eq:IC}) under normal errors. }
First, for most rank combinations, the joint selection probabilities approach one as $T$ increases for a fixed $N$. Selection is more difficult when both true ranks are larger, and the improvement with $T$ is not always monotone. Second, larger cross-sectional dimensions can reduce selection accuracy in smaller samples. Comparing Case 2 with Case 1, however, the presence of cross-sectional dependence and heteroskedasticity does not systematically reduce selection accuracy.

Additional results under standardized $t(5)$ and mixture-normal errors are reported in Section~S5.1 of the supplementary material, while Section~S5.5 examines sensitivity to the penalty constant. Overall, the simulations indicate generally high rank-selection accuracy, with finite-sample performance depending on the sample dimensions, true ranks, and penalty choice.

\begin{table}[h!]
\caption{
Estimation accuracy of $\widehat{r}_g$ and $\widehat{r}_l$ for Cases 1 and 2 under normal errors.
}
\vspace{-0.1cm}
\label{rank_norm}
{\centering
\footnotesize{
\setlength{\tabcolsep}{2pt}
\renewcommand{\arraystretch}{1}
\begin{tabular}{cc*{4}{ccc}}
\toprule
& & \multicolumn{3}{c}{$r_g=r_l=1$} & \multicolumn{3}{c}{$r_g=1,\ r_l=2$} & \multicolumn{3}{c}{$r_g=2,\ r_l=1$} & \multicolumn{3}{c}{$r_g=r_l=2$} \\
\cmidrule(lr){3-5}\cmidrule(lr){6-8}\cmidrule(lr){9-11}\cmidrule(lr){12-14}
$N$ & $T$ & $\widehat r_g$ & $\widehat r_l$ & $(\widehat r_g,\widehat r_l)$ & $\widehat r_g$ & $\widehat r_l$ & $(\widehat r_g,\widehat r_l)$ & $\widehat r_g$ & $\widehat r_l$ & $(\widehat r_g,\widehat r_l)$ & $\widehat r_g$ & $\widehat r_l$ & $(\widehat r_g,\widehat r_l)$ \\ \hline
\multicolumn{14}{c}{Panel A: Case 1} \\ \hline
60 & 1000 & 1.00 & 0.98 & 0.98 & 0.99 & 0.99 & 0.98 & 0.95 & 0.93 & 0.90 & 0.96 & 0.91 & 0.87 \\
 & 1500 & 1.00 & 1.00 & 1.00 & 1.00 & 0.99 & 0.99 & 0.96 & 1.00 & 0.96 & 0.91 & 0.89 & 0.80 \\
 & 2000 & 1.00 & 1.00 & 1.00 & 1.00 & 1.00 & 1.00 & 0.98 & 1.00 & 0.98 & 0.98 & 0.94 & 0.92 \\
90 & 1000 & 0.91 & 0.86 & 0.79 & 0.88 & 1.00 & 0.88 & 1.00 & 0.89 & 0.89 & 0.97 & 0.97 & 0.94 \\
 & 1500 & 1.00 & 1.00 & 1.00 & 1.00 & 1.00 & 1.00 & 1.00 & 1.00 & 1.00 & 1.00 & 0.99 & 0.99 \\
 & 2000 & 1.00 & 1.00 & 1.00 & 1.00 & 1.00 & 1.00 & 1.00 & 1.00 & 1.00 & 1.00 & 1.00 & 1.00 \\
120 & 1000 & 0.81 & 0.68 & 0.54 & 0.79 & 1.00 & 0.79 & 0.94 & 0.61 & 0.57 & 0.92 & 0.99 & 0.91 \\
 & 1500 & 1.00 & 0.99 & 0.99 & 1.00 & 0.99 & 0.99 & 1.00 & 1.00 & 1.00 & 0.98 & 1.00 & 0.98 \\
 & 2000 & 1.00 & 1.00 & 1.00 & 1.00 & 0.98 & 0.98 & 1.00 & 1.00 & 1.00 & 0.95 & 1.00 & 0.95 \\
\hline
\multicolumn{14}{c}{Panel B: Case 2} \\ \hline
60 & 1000 & 0.99 & 0.98 & 0.97 & 0.99 & 0.98 & 0.97 & 0.89 & 0.99 & 0.88 & 0.89 & 0.94 & 0.83 \\
 & 1500 & 1.00 & 1.00 & 1.00 & 1.00 & 0.99 & 0.99 & 0.93 & 1.00 & 0.93 & 0.91 & 0.91 & 0.82 \\
 & 2000 & 1.00 & 1.00 & 1.00 & 1.00 & 0.99 & 0.99 & 0.95 & 1.00 & 0.95 & 0.92 & 0.93 & 0.85 \\
90 & 1000 & 0.96 & 0.88 & 0.85 & 0.94 & 1.00 & 0.94 & 0.97 & 0.87 & 0.85 & 0.99 & 0.99 & 0.98 \\
 & 1500 & 1.00 & 0.99 & 0.99 & 1.00 & 1.00 & 1.00 & 1.00 & 0.99 & 0.99 & 0.96 & 0.97 & 0.93 \\
 & 2000 & 1.00 & 1.00 & 1.00 & 1.00 & 1.00 & 1.00 & 1.00 & 1.00 & 1.00 & 0.98 & 0.98 & 0.96 \\
120 & 1000 & 0.81 & 0.64 & 0.55 & 0.77 & 0.97 & 0.75 & 0.96 & 0.61 & 0.58 & 0.97 & 0.99 & 0.96 \\
 & 1500 & 1.00 & 0.99 & 0.99 & 1.00 & 1.00 & 1.00 & 1.00 & 1.00 & 1.00 & 0.94 & 0.98 & 0.92 \\
 & 2000 & 1.00 & 1.00 & 1.00 & 1.00 & 0.99 & 0.99 & 1.00 & 1.00 & 1.00 & 1.00 & 0.98 & 0.98 \\
\bottomrule
\end{tabular}}
\par}
\vspace{0.12cm}
\footnotesize{Notes:
This table reports the proportions of correct rank selection for Cases 1 and 2 under normal errors. The true global ($r_g$) and local ($r_l$) ranks are set to $\{1,2\}$, yielding four combinations shown in the column groups.
Columns headed $\widehat r_g$ and $\widehat r_l$ report the proportions of correctly selecting the global and local ranks, respectively, while $(\widehat r_g,\widehat r_l)$ reports the proportion of jointly selecting both ranks correctly. Panels A and B correspond to Cases 1 and 2, respectively.}
\end{table}

\section{Empirical Applications}\label{sec4}

In this section, we apply the HGFM to returns on S\&P 100 constituents from five industry groups. We first estimate the model parameters using the QMLE method. We then compare explanatory fit and investment performance with competing models and examine transaction costs and alternative empirical specifications.

\subsection{Data Description}\label{sec:emp_data}

We analyze daily returns from the Center for Research in Security Prices (CRSP) for S\&P~100 constituents from January 1, 2016, to December 31, 2025. After excluding stocks with insufficient return histories, we obtain a balanced panel of 98 stocks with 2,514 daily observations per stock. The stocks are assigned to five industry groups based on the Global Industry Classification Standard (GICS), with group membership held fixed throughout the sample. Table~S12 in the supplementary material reports the group composition and complete stock list. We construct the global and local portfolio weights using market beta, market equity, and book-to-market, each ranked across the sample stocks and scaled to $[-1,1]$. Local factors use these standardized characteristics within each industry group.

\subsection{Empirical Findings}

\subsubsection{Global and Local Factor Model Performance}\label{sec:emp_models}

The information criterion in~\eqref{eq:IC} selects $\widehat r_g=\widehat r_l=1$, giving one global factor and one local factor for each industry group. We estimate this specification by QMLE using Algorithm~\ref{al1} with three random initializations. The global factor is denoted by G1, and the five local factors by L1--L5, following the industry order in Table~S12.
We compare the full HGFM with the capital asset pricing model (CAPM), the Fama--French three-factor model (FF3) \citep{Fama1993Common}, and the Fama--French five-factor model (FF5) \citep{fama2015five}. We also include equal- and value-weighted portfolios of the sample stocks, denoted by EW98 and VW98, respectively. To assess the contributions of the global and local factors, we consider two subsets of the jointly estimated factors: HGFM (global), containing G1, and HGFM (local), containing L1--L5.

\paragraph{Explanatory fit.}

We assess explanatory performance using Total $R^2$ and cross-sectional $R^2$ (CS $R^2$) \citep{kelly2019characteristics,feng2024deep}. Total $R^2$ measures the fit of realized excess returns across assets and dates, whereas CS $R^2$ measures the fit of average excess returns across assets. Specifically,
\begingroup\setlength{\abovedisplayskip}{4pt}\setlength{\belowdisplayskip}{4pt}\setlength{\abovedisplayshortskip}{2pt}\setlength{\belowdisplayshortskip}{2pt}
\begin{align*}
\mathrm{Total}\ R^2
=1-\frac{\sum_{i=1}^{N}\sum_{t=1}^{T}(Y^e_{it}-\widehat Y^e_{it})^2}
{\sum_{i=1}^{N}\sum_{t=1}^{T}(Y^e_{it})^2}, 
\mathrm{CS}\ R^2
=1-\frac{\sum_{i=1}^{N}(\overline Y^e_i-\overline{\widehat Y}^e_i)^2}
{\sum_{i=1}^{N}(\overline Y^e_i)^2},
\end{align*}
\endgroup
where $Y^e_{it}=Y_{it}-r_{f,t}$, $Y_{it}$ is the $i$th element of $\mathbf Y_t$, and $r_{f,t}$ is the risk-free return. The quantity $\widehat Y^e_{it}$ denotes the fitted excess return from each model. The quantities $\overline Y^e_i$ and $\overline{\widehat Y}^e_i$ denote the time averages of observed and fitted excess returns, respectively. 

Table~\ref{pricing_combined} reports Total $R^2$ for the full stock panel and each industry group. {We find that the full model has the highest Total $R^2$ among all models considered, both for the full panel and within each industry group.} For the full panel, its Total $R^2$ is 0.538, compared with 0.362 for HGFM (global), 0.524 for HGFM (local), and 0.435 for FF5, the best-performing benchmark. The local factors account for much of the explanatory fit, while adding the global factor provides further gains across all five industries. These results indicate that the global and local factors together explain realized return variation more effectively than either subset alone or the benchmark models.

Table~\ref{pricing_cs} reports the corresponding results for CS $R^2$. For the full panel, HGFM achieves 0.836, exceeding all benchmark models but remaining slightly below HGFM (local). {At the industry level, our model outperforms all benchmarks in technology and communication (I1), but not in the other groups.} Overall, the benefits of combining global and local factors are particularly evident in explaining realized returns, while improvements in average-return fit vary across industry groups.

\begin{table}[h!]
\caption{Total $R^2$ for the full stock panel and industry groups.}
\label{pricing_combined}
\vspace{-0.1cm}
{\centering
{\footnotesize\setlength{\tabcolsep}{8pt}\renewcommand{\arraystretch}{1}
\begin{tabular}{lcccccc}
\toprule
Model & All & I1 & I2 & I3 & I4 & I5 \\
\midrule
\multicolumn{7}{c}{\underline{Panel A: HGFM}} \\
HGFM & 0.538 & 0.500 & 0.425 & 0.428 & 0.708 & 0.580 \\
HGFM (global) & 0.362 & 0.323 & 0.305 & 0.241 & 0.540 & 0.362 \\
HGFM (local) & 0.524 & 0.481 & 0.396 & 0.427 & 0.701 & 0.571 \\
\midrule
\multicolumn{7}{c}{\underline{Panel B: Benchmarks}} \\
CAPM & 0.344 & 0.369 & 0.301 & 0.204 & 0.480 & 0.310 \\
FF3 & 0.425 & 0.413 & 0.347 & 0.222 & 0.634 & 0.427 \\
FF5 & 0.435 & 0.420 & 0.363 & 0.230 & 0.646 & 0.436 \\
EW98 & 0.368 & 0.312 & 0.306 & 0.233 & 0.553 & 0.391 \\
VW98 & 0.332 & 0.374 & 0.295 & 0.207 & 0.451 & 0.284 \\
\bottomrule
\end{tabular}
}
\par}
\vspace{0.12cm}
\footnotesize{Notes: This table reports Total $R^2$ for the full stock panel (All) and the five industry groups (I1--I5). Panel A presents results for the full HGFM and its global-only and local-only subsets, using factors from the joint estimation. Panel B presents results for CAPM, FF3, FF5, and the equal- and value-weighted sample portfolios (EW98 and VW98).
}

\end{table}

\begin{table}[h!]
\caption{CS $R^2$ for the full stock panel and industry groups.}
\label{pricing_cs}
\vspace{-0.1cm}
{\centering
{\footnotesize\setlength{\tabcolsep}{8pt}\renewcommand{\arraystretch}{1}
\begin{tabular}{lcccccc}
\toprule
Model & All & I1 & I2 & I3 & I4 & I5 \\
\midrule
\multicolumn{7}{c}{\underline{Panel A: HGFM}} \\
HGFM & 0.836 & 0.895 & 0.712 & 0.690 & 0.919 & 0.800 \\
HGFM (global) & 0.740 & 0.666 & 0.697 & 0.676 & 0.923 & 0.841 \\
HGFM (local) & 0.839 & 0.851 & 0.798 & 0.692 & 0.916 & 0.849 \\
\midrule
\multicolumn{7}{c}{\underline{Panel B: Benchmarks}} \\
CAPM & 0.768 & 0.707 & 0.738 & 0.660 & 0.928 & 0.866 \\
FF3 & 0.772 & 0.707 & 0.718 & 0.703 & 0.939 & 0.883 \\
FF5 & 0.767 & 0.707 & 0.683 & 0.712 & 0.937 & 0.889 \\
EW98 & 0.732 & 0.676 & 0.691 & 0.670 & 0.905 & 0.785 \\
VW98 & 0.793 & 0.752 & 0.768 & 0.690 & 0.927 & 0.856 \\
\bottomrule
\end{tabular}}
\par}
\vspace{0.12cm}
\footnotesize{Notes: This table reports CS $R^2$ for the full stock panel (All) and the five industry groups (I1--I5). 
The table format follows Table \ref{pricing_combined}. 
}
\end{table}

\paragraph{Investment performance.}

We evaluate investment performance using mean-variance efficient (MVE) portfolios formed from the factors. For multifactor models, portfolio weights are determined using the means and covariance matrix of daily factor excess returns. Single-factor models use the factor directly.
Table~\ref{performance_investment} reports investment performance before transaction costs. The full HGFM achieves the highest annualized Sharpe ratio among the models considered, at 1.32, compared with 0.84 for HGFM (global), 1.09 for HGFM (local), and 0.98 for FF3. It also has the lowest volatility and maximum drawdown, despite a lower mean return than either factor subset. These results suggest that combining global and local factors improves the balance between return and risk within the evaluation sample.

We further examine transaction costs and alternative industry groupings in Tables~S13 and~S14 of the supplementary material. HGFM retains a higher Sharpe ratio than the equal- and value-weighted sample portfolios at all reported transaction cost rates. The alternative grouping provides an additional check on the specification, showing that explanatory fit and investment performance depend on how industries are grouped.

\begin{table}[h!]
\caption{Investment performance.}
\label{performance_investment}
\vspace{-0.1cm}
{\centering
{\footnotesize\setlength{\tabcolsep}{8pt}\renewcommand{\arraystretch}{1}
\begin{tabular}{lcccc}
\toprule
Model & AVG (\%) & SD (\%) & SR & MDD (\%) \\
\midrule
\multicolumn{5}{c}{\underline{Panel A: HGFM}} \\
HGFM & 0.75 & 1.48 & 1.32 & 8.12 \\
HGFM (global) & 1.22 & 4.24 & 0.84 & 20.20 \\
HGFM (local) & 1.71 & 4.86 & 1.09 & 28.94 \\
\midrule
\multicolumn{5}{c}{\underline{Panel B: Benchmarks}} \\
CAPM & 1.26 & 4.54 & 0.82 & 24.90 \\
FF3 & 0.84 & 2.32 & 0.98 & 12.23 \\
FF5 & 0.67 & 1.74 & 0.97 & 9.75 \\
EW98 & 1.32 & 4.48 & 0.87 & 21.56 \\
VW98 & 1.38 & 4.43 & 0.93 & 23.64 \\
\bottomrule
\end{tabular}}
\par}
\vspace{0.12cm}
\footnotesize{Notes: This table reports investment performance across different models. AVG and SD are the mean and standard deviation of monthly total returns (\%). SR is the Sharpe ratio of monthly excess returns, annualized by $\sqrt{12}$. MDD is the maximum drawdown based on monthly returns (\%).
}
\end{table}

\subsubsection{Economic Interpretation}\label{sec:emp_interpretation}

We interpret the factor portfolios through the estimated characteristic coefficients that determine their stock weights. We report $\widehat{\mathbf C}_g$ and $\widehat{\mathbf C}_{l,s}$ after the weighted SVD normalization in Proposition~\ref{prop:svd_internal_transformation}.
{The entries of $\widehat{\mathbf C}_g$ and $\widehat{\mathbf C}_{l,s}$ are the estimated coefficients used to form portfolio weights, rather than coefficients from a separate regression of factors on characteristics. The weights are $\widehat{\mathbf C}_g\mathbf Z_{t-1}$ and $\widehat{\mathbf C}_{l,s}\mathbf Z_{s,t-1}$ before rescaling each portfolio so that its absolute stock weights sum to one. A positive coefficient means that, for two stocks with the same values of the other characteristics, the stock with the higher value of that characteristic receives a higher weight before this rescaling. These coefficients describe portfolio composition, not risk premia.}

\begin{table}[h!]
\caption{Estimated characteristic coefficients of global and local factors.}
\label{global_characteristic_coefficients}
\vspace{-0.1cm}
{\centering
{\footnotesize
\setlength{\tabcolsep}{0.6em}
\renewcommand{\arraystretch}{1}
\begin{tabular}{lccc}
\toprule
Factor & Market beta & Market equity & Book-to-market \\
\midrule
G1 & -0.014 & 0.127 & 0.046 \\
\midrule
L1 & 0.208 & 0.150 & -0.235 \\
L2 & -0.339 & 0.320 & 0.243 \\
L3 & 0.065 & 0.226 & 0.112 \\
L4 & 0.032 & -0.030 & 0.265 \\
L5 & 0.128 & -0.050 & 0.225 \\
\bottomrule
\end{tabular}
}
\par}
\vspace{0.12cm}
\footnotesize{Notes: This table reports the estimated coefficients on market beta, market equity, and book-to-market used to construct the global and local portfolio weights.
G1 denotes the global factor, and L1--L5 denote the local factors. 
}
\end{table}

{Table~\ref{global_characteristic_coefficients} reports the entries of $\widehat{\mathbf C}_g$ and $\widehat{\mathbf C}_{l,s}$ corresponding to market beta, market equity, and book-to-market.} Several patterns emerge from these estimates. 
First, the global factor assigns greater weight to larger stocks and stocks with higher book-to-market ratios, while its market-beta coefficient is small and negative. Second, local factors differ across industries. In technology and communication (I1), L1 favors larger, higher-beta stocks with lower book-to-market ratios, whereas L2 in the consumer group (I2) favors larger, lower-beta stocks with higher book-to-market ratios. Third, market equity has the largest coefficient for healthcare (I3), while book-to-market has the largest coefficient for financials (I4) and industrials, energy, and utilities (I5). L5 also assigns greater weight to higher-beta and smaller stocks. These patterns illustrate how observable characteristics shape the global and local portfolio weights. {Section~S6.4 of the supplementary material reports stock weights and gross-exposure shares for these factor portfolios.}

\section{Conclusion} \label{sec5}

This paper proposes a Hierarchical Group Factor Model (HGFM) with global and local factors constructed from observed returns using characteristic-driven weights. Conditional on identification of the time-varying interaction matrix, cross-group characteristic variation separates the global and local coefficient matrices. Normalization using the empirical second-moment matrices of the observed characteristics identifies their factor representations up to paired sign changes. {Under a fixed characteristic design and the stated conditions, we establish local consistency for the coefficient and log-variance estimators, as well as consistency and fixed-block asymptotic normality for the normalized factor parameters.} We also propose an information criterion that consistently selects the global and common local ranks.

In the application to S\&P~100 constituents, {the global and local factor portfolios place different weights on stock characteristics.} Combining these factors improves realized-return fit across all industry groups relative to either subset alone and the benchmarks, while delivering the highest Sharpe ratio and lowest volatility and maximum drawdown among the portfolios considered. {The combined portfolio's Sharpe ratio remains above those of equal- and value-weighted sample portfolios after transaction costs, although performance varies with industry grouping.}

We suggest four directions for future research. 
First, the model can be extended to incorporate exogenous predictors and
autoregressive dynamics, thereby allowing it to capture observed covariate effects and temporal dependence. Second, the framework could be extended to allow the number of characteristics to grow, subject to further identification conditions and asymptotic analysis. Third, this paper focuses on grouped vector-valued observations. Related hierarchical structures could be considered for matrix-valued time series \citep[e.g.,][]{wang2019factor,zhang2025multilevel}, three-level factor structures \citep[e.g.,][]{jin2025three}, and tensor factor models \citep[e.g.,][]{chen2022factor,chen2024semi}. Finally, the model could be used to examine global and local components within broader collections of candidate factors \citep{jensen2023there} and incorporate machine-learning-based factor construction.

\paragraph{Data Availability Statement} The data that support the findings of this study are available from Wharton Research Data Services. Restrictions apply to the availability of these data, which were used under license for this study.

\bibliographystyle{chicago}
{\setstretch{1.15}\setlength{\bibsep}{0pt}\bibliography{ref}}
\end{document}